\documentclass[aps,prd,twocolumn,nofootinbib]{revtex4}
\usepackage{graphicx}
\usepackage[utf8]{inputenc}
\usepackage{placeins}
\usepackage{amssymb, amsmath}
\usepackage[normalem]{ulem}
\usepackage{tabularx}
\newcommand{\Ms}{M_{\odot}}
\newcommand{\ns}{n_{\rm sat}}
\newcommand{\npe}{$n$-$p$-$e$~}

\usepackage[dvipsnames, usenames]{xcolor}

\begin{document}

\title{Impact of the nuclear symmetry energy on the post-merger phase of a binary neutron star coalescence}

\author{Elias R. Most}
\email{emost@princeton.edu}
\author{Carolyn A. Raithel}
\email{craithel@ias.edu}
\affiliation{School of Natural Sciences, Institute for Advanced Study, 1 Einstein Drive, Princeton, NJ 08540, USA}
\affiliation{Princeton Center for Theoretical Science, Jadwin Hall, Princeton University, Princeton, NJ 08544, USA}
\affiliation{Princeton Gravity Initiative, Jadwin Hall, Princeton University, Princeton, NJ 08544, USA}
\thanks{Both authors have contributed equally to this work.}
\date{May 2021}

\begin{abstract}
The nuclear symmetry energy plays a key role in determining the equation of state of dense, neutron-rich matter, which governs the properties of both terrestrial nuclear matter as well as astrophysical neutron stars. A recent measurement of the neutron skin thickness from the PREX collaboration has lead to new constraints on the slope of the nuclear symmetry energy, $L$, which can be directly compared to inferences from gravitational-wave observations of the first binary neutron star merger inspiral, GW170817 
In this paper, we explore a new regime for potentially constraining the slope, $L$, of the nuclear symmetry energy with future gravitational wave events: the post-merger phase a binary neutron star coalescence.  In particular, we go beyond the inspiral phase, where imprints of the slope
parameter $L$ may be inferred from measurements of the tidal
deformability, to consider imprints on the post-merger dynamics,
 gravitational wave emission, and dynamical mass ejection. 
 To this end, we perform a set of targeted neutron star merger
simulations in full general relativity using new finite-temperature 
equations of state, which systematically vary $L$.
We find that the post-merger dynamics and gravitational wave emission are mostly insensitive to the slope of the nuclear symmetry energy.
In contrast, we find that dynamical mass ejection contains a weak imprint of $L$, with large values of $L$ leading to systematically enhanced ejecta.  
\end{abstract}

\maketitle

\section{Introduction}

Determining the equation of state (EoS) of dense, neutron-rich matter is a common goal in both modern nuclear physics and astrophysics. One of the key ingredients to the neutron-rich EoS is the nuclear symmetry energy, which characterizes the difference in energy between symmetric nuclear matter and pure neutron matter. The symmetry energy is often represented as a series expansion in density, with leading-order coefficients related to the value of the symmetry energy at the nuclear saturation density, $S$, and its slope, $L$, according to
\begin{equation}
\label{eq:Esym}
E_{\rm sym}(n) = S + \frac{L}{3} \left( \frac{n}{\ns}-1 \right) + \mathcal{O}\left[\left( \frac{n}{\ns}-1 \right)\right]^2
\end{equation}
where $\ns=0.16$~fm$^{-3}$ is the nuclear saturation density  \cite{Piekarewicz:2008nh}.

A wide range of experimental efforts have placed constraints on $S$ and $L$, including from fits to nuclear masses, measurements of the giant dipole resonance and electric dipole polarizability of $^{208}$Pb, and observations of isospin diffusion or multifragmentation in heavy ion collisions \citep{Tsang:2012se,Lattimer:2012xj,Oertel:2016bki}. Recently, the Lead Radius Experiment (PREX-II) reported new constraints on the neutron radius of $^{208}$Pb which, when combined with results from the original PREX-I experiment \cite{Abrahamyan:2012gp,Horowitz:2012tj}, imply a neutron skin thickness of $R_{\rm skin}^{^{208}\rm{Pb}} = 0.283\pm0.071$~fm \cite{PREX:2021umo}. From this measurement, Ref. \cite{Reed:2021nqk} constrained the slope of the symmetry energy to $L=106\pm37$~MeV, which is larger than many previous constraints from microscopic calculations or experimental measurements \cite{Tsang:2012se,Lattimer:2012xj,Oertel:2016bki,Li:2021thg}.

\begin{figure}[!ht]
\centering
\includegraphics[width=0.5\textwidth]{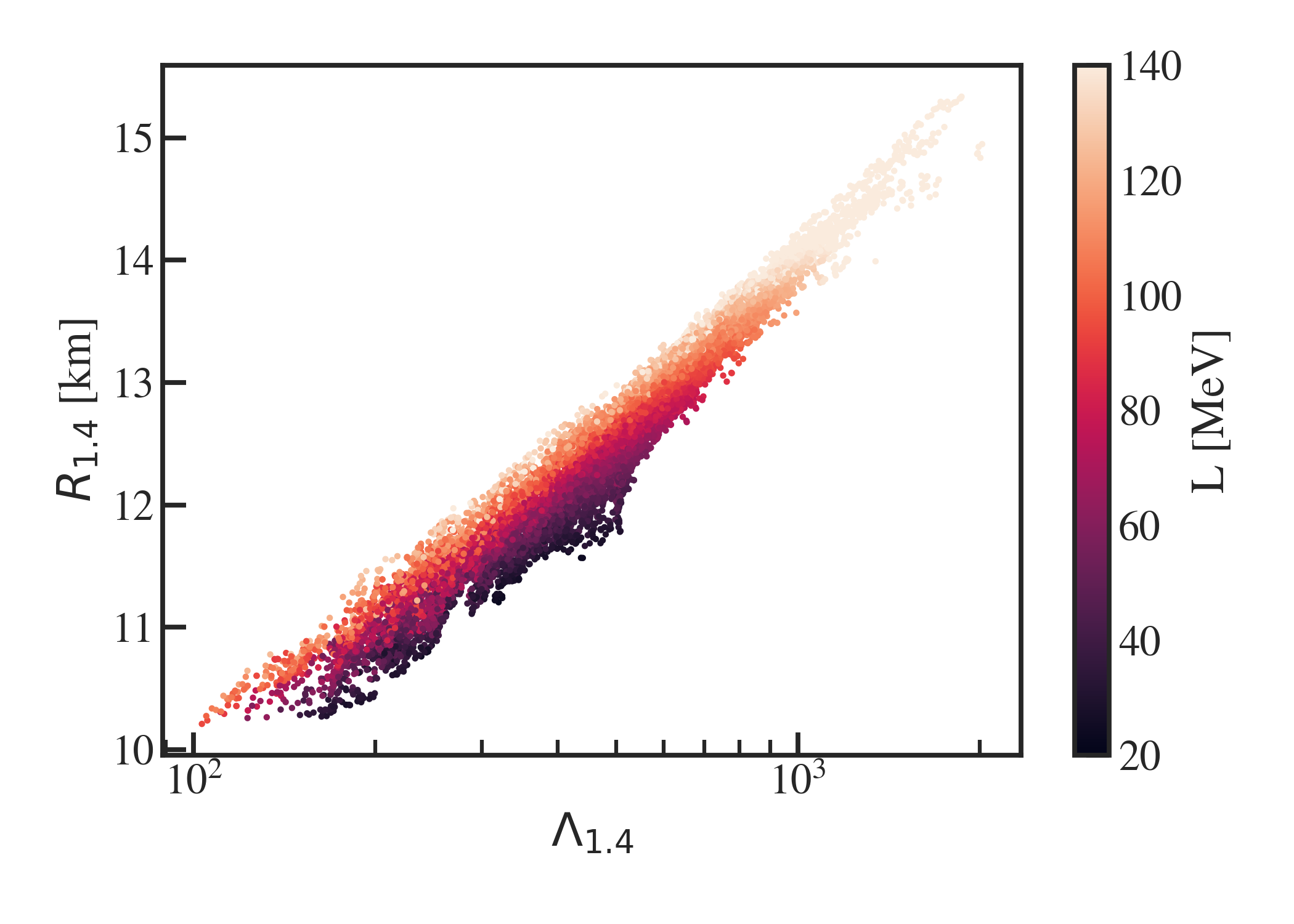}
\caption{\label{fig:RLambdaL} Correlations between $R_{1.4}$ and $\Lambda_{1.4}$ for a sample of $>10,000$
piecewise polytropic EoSs. Each EoS has five polytropic segments, spaced log-uniformly in density between 0.5 and 7.4$\ns$, with uniformly-drawn pressures. All EoSs support a maximum mass of at least 1.97~$\Ms$. The symmetry energy slope is extracted according to eq.~(\ref{eq:LofP}). We find that, although $R_{1.4}$ and $\Lambda_{1.4}$ are well correlated, there is a significant scatter in the trend, which depends approximately monotonically on the value of $L$.}
\end{figure}

Following this new measurement of $L$, several studies have recently investigated its impact on the neutron star EoS.
Neutron stars, which contain neutron-rich matter and probe densities around and above the nuclear saturation density, are an ideal laboratory for studying the symmetry energy. It has long been known that the slope of the symmetry energy correlates strongly with neutron star radius (\cite{Lattimer:2000nx}, see also Fig.~\ref{fig:RLambdaL}). The radius in turn correlates with the neutron star tidal deformability \cite{Yagi:2013awa,Yagi:2015pkc,De:2018uhw,Raithel:2018ncd}. Perhaps not surprisingly then, $L$ can also affect the gravitational wave emission during a binary neutron star inspiral (e.g., \cite{Fattoyev:2013rga}). Measurements of these astrophysical quantities can thus, in principle, provide constraints on $L$ that are complementary to those inferred from laboratory-based experiments.
 
To illustrate this behavior, we show the inter-correlations between $R_{1.4}$, $\Lambda_{1.4}$, and the slope of the symmetry energy in Fig.~\ref{fig:RLambdaL}, where $R_{1.4}$ and $\Lambda_{1.4}$ are the characteristic radius and tidal deformability of a 1.4~$\Ms$ neutron star, respectively. Figure~\ref{fig:RLambdaL} was generated from a large sample of piecewise polytropic EoSs, which were constructed to uniformly sample the pressure at densities above half of the nuclear saturation density (see Sec.~\ref{sec:EOS} and Ref. \cite{Raithel:2016bux} for further details). Figure~\ref{fig:RLambdaL} shows that, although $R_{1.4}$ and $\Lambda_{1.4}$ are indeed well correlated, there is significant scatter in the relationship, which depends sensitively on the value of $L$ in an approximately monotonic fashion. For example, for fixed values of $\Lambda_{1.4}\lesssim500$, the corresponding value of $R_{1.4}$ can vary by nearly a kilometer, with $L$ likewise varying from $\lesssim 40$~MeV to more than 100~MeV. While these correlations hold for the radii of intermediate mass stars, the dependence on $L$ breaks down for higher-mass stars ($M>1.8~\Ms$), which are governed less strongly governed by saturation physics \cite{Alam:2016cli}.

Taking advantage of the types of correlations shown in Fig.~\ref{fig:RLambdaL}, as well as correlations between $\Lambda_{1.4}$ and the higher-order symmetry energy coefficients, many studies have used the measurement of the tidal deformability from GW170817 \cite{LIGOScientific:2017vwq,LIGOScientific:2018hze} to derive new constraints on the nuclear symmetry energy \cite{Krastev:2018nwr,Malik:2018zcf,Carson:2018xri,Raithel:2019ejc, Zhang:2018bwq,Tsang:2019jva}. {For example, Ref.~\cite{Raithel:2019ejc} demonstrated that, for a mono-parametric family of EoSs, GW170817 implies small values of $9\lesssim L \lesssim 65$~MeV}. In a recent study combining astrophysical data from GW170817, NICER, and the existence of massive pulsars, together with theoretical constraints from chiral effective field theory, Ref.~\cite{Essick:2021kjb} confirmed that the astrophysically-inferred slope of the symmetry energy ($L=52\substack{+20\\-18}$~MeV) is in mild tension with the PREX-II result. For a recent review on the status of astrophysical and laboratory constraints on the symmetry energy in light of GW170817 and the PREX-II experiment, see \cite{Li:2021thg}.

In this work, we explore a new regime for probing the nuclear symmetry energy:  the post-merger phase of a binary neutron star merger. The post-merger phase probes higher densities and larger temperatures ($T>40\, \rm MeV$) than in the inspiral.
As a result, astrophysical observables of the post-merger phase provide an ideal laboratory for probing the properties of hot, dense matter.
For example, the gravitational waves (GWs) emitted by the post-merger remnant are expected to be a sensitive probe of the underlying cold EoS \cite{Baiotti:2016qnr,Paschalidis:2016vmz,Bauswein:2019ybt,Bernuzzi:2020tgt,Radice:2020ddv}, with additional corrections from the finite-temperature part of the EoS \cite{Bauswein:2010dn,Raithel:2021hye}.
The dynamics of the post-merger phase will also influence the quantity, velocity, and composition of the dynamical ejecta , which can in turn influence the associated kilonova, if detectable \cite{Metzger:2019zeh,Radice:2018pdn,Kawaguchi:2019nju,Shibata:2019wef}. Modeling all of these effects requires accurate numerical relativity simulations of the post-merger phase that account for all relevant physical processes, including weak-interactions and finite temperature effects, see e.g. \cite{Radice:2020ddv} for a recent review.

Whereas most early studies resorted to ideal-fluid descriptions for the finite temperature part of the EoS and neglected nuclear composition entirely, many recent studies have instead made use of a limited number of publicly available EoS tables\footnote{{E.g., from the CompOSE database, \texttt{https://compose.obspm.fr/}, or from \texttt{https://stellarcollapse.org/.}}} that enable self-consistent finite-temperature effects and out-of-(weak) equilibrium composition effects. One drawback of the latter approach, however, is the relatively small sample of available models, which can vary from one another in multiple nuclear parameters simultaneously, rendering systematic studies of the impact of individual nuclear parameters on merger simulations nearly impossible.

In this work, we take a third approach, enabled by a recently-developed framework for extending arbitrary cold EoSs to finite-temperatures and arbitrary compositions. The finite-temperature part of this EoS framework utilizes a Fermi Liquid Theory approach for including the leading-order effects of degeneracy, while the extrapolation to non-equilibrium compositions is based on a parametrization of the nuclear symmetry energy \cite{Raithel:2019gws}. 
The ability of this framework to model finite-temperature effects in neutron star merger simulations was recently explored by Ref. \cite{Raithel:2021hye}.
In summary, this framework allows for the construction of new \textit{parametric, finite-temperature} EoSs, which are ideally suited for systematic investigations of EoS imprints in neutron star mergers. 

In particular, many previous studies have found that the post-merger GW emission or mass ejecta depend on the characteristic radius or tidal deformability of the underlying EoS \cite{Baiotti:2016qnr,Paschalidis:2016vmz,Radice:2018pdn,Bauswein:2019ybt,Shibata:2019wef,Bernuzzi:2020tgt,Radice:2020ddv}. These interconnected dependencies typically would make it difficult to disentangle variations in these macroscopic properties from any variations in the nuclear model, e.g., in $L$.
To start to resolve this problem, in this work, we explicitly construct a set of seven new EoS models that fix $R_{1.4}$ or $\Lambda_{1.4}$, while systematically varying in $L$.
In this way, we aim to disentangle the role of these macroscopic properties from the possible role of $L$ in determining the post-merger observables.

To further restrict the comparison of the EOSs to the supranuclear part, where $L$ plays a role, we construct each EoS to have an identical finite-temperature component and to follow the same tabulated EoS at densities below half of the nuclear saturation density (SFHo, \cite{Steiner:2012rk}). In this way, we ensure that the only difference in these new EoSs is in the cold physics at supranuclear densities, while also capturing low density effects, such as the formation of bound nuclei, by using a tabulated nuclear model.

In order to study the role of $L$ in the post-merger phase, we perform numerical simulations of binary neutron star mergers in full general relativity using each of these new EoSs. Our simulations follow the last few orbits of the binary inspiral, and continue through the merger and until $\sim$25~ms post-merger. We use these simulations to explore, in particular, the post-merger dynamics, post-merger gravitational wave emission, and dynamical mass ejection.
We find that the slope $L$ of the symmetry energy does not leave clear imprints in the post-merger dynamics or GW emission, but rather find that these processes depend more sensitively on the high-density part of the EoS. 
In contrast, we find some first indication that the amount of dynamically ejected material correlates with the slope of the symmetry energy, with large values of $L$ leading to the production of significantly more dynamical ejecta. 
These links between $L$ and the ejecta may affect aspects of the electromagnetic counterpart to the merger, such as an X-ray rebrightening \cite{Hotokezaka:2018gmo}, as has recently been observed for GW170817 \cite{Hajela:2021faz,Balasubramanian:2021kny}.

The outline of the paper is as follows. We start in Sec.~\ref{sec:EOS} by describing the construction of the EoSs used in this work.  We describe the numerical setup of our simulations in Sec.~\ref{sec:numerics} and the initial conditions in Sec.~\ref{sec:id}. In Sec.~\ref{sec:results}, we present the results of our merger simulations, discussing the properties of the post-merger remnant in Sec.~\ref{sec:remnant}, the dynamical ejecta in Sec.~\ref{sec:ejecta}, and the gravitational wave content in Sec.~\ref{sec:GW}. \\

Unless explicitly stated, we adopt units of $c=G=k_B=1$.

\section{Methods}

In the following, we give a detailed overview on the construction of the EoSs used 
in this work. We also briefly summarize the numerical methods and setup used to perform our simulations.
\subsection{Equations of state}

\label{sec:EOS}
In order to explore the impact of $L$ on the post-merger phase of a binary neutron star coalescence, we construct a set of seven new EoS tables, which systematically vary the slope $L$ of the symmetry energy while keeping particular macroscopic properties fixed. In this section, we summarize the framework used to construct these EoSs, starting with a brief overview of our approach. 

For all models, we start with an identical, finite-temperature EoS table (SFHo, \cite{2013ApJ...774...17S})\footnote{The SFHo table was provided by \texttt{stellarcollapse.org}.}, which we use to describe the matter at densities up to half the nuclear saturation density, $\ns$. At these low densities, the SFHo EoS table uses the statistical model of \cite{Hempel:2009mc} to describe the non-uniform (i.e., bound) matter in nuclear statistical equilibrium, while the unbound nucleons are described by the SFHo model for relativistic mean field interactions. At densities above $0.5\ns$, we switch to a piecewise polytropic framework to represent the EoS of cold, dense matter in $\beta$-equilibrium. This choice provides us with maximum flexibility for exploring a wide region of the zero-temperature EoS parameter space. We then use the framework of \cite{Raithel:2019gws} to extrapolate the cold, $\beta$-equilibrium EoS to finite temperatures and arbitrary electron fractions. Throughout this paper, we limit our consideration to neutron-proton-electron matter. Crucially, the low-density physics and the finite-temperature part of the EoSs are held constant between all models, in order to ensure a systematic comparison of the cold supranuclear regime and, hence, of $L$.

\begin{figure}
\centering
\includegraphics[width=0.425\textwidth]{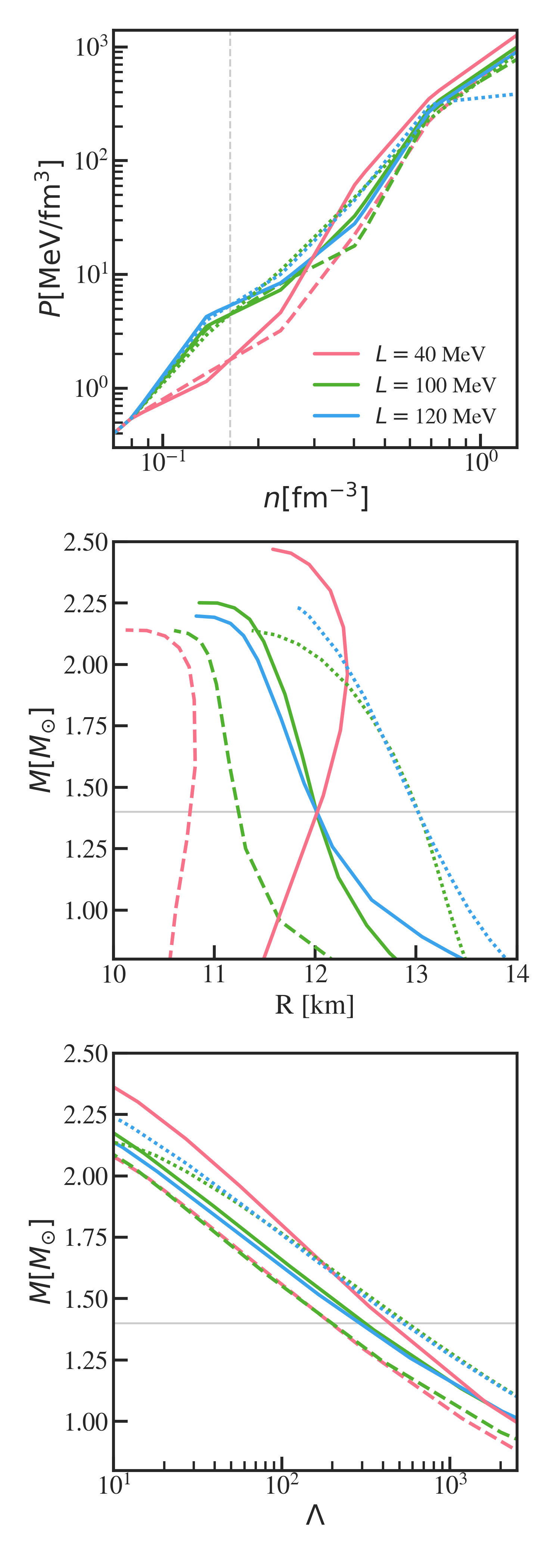}
\caption{\label{fig:MR} Equations of state included in our sample. The top panel shows the zero-temperature, $\beta$-equilibrium pressure; the middle panel shows the mass-radius relation; and the bottom panel shows the tidal deformabilities. The dashed lines indicate EoSs with $R_{1.4}\simeq11$~km, which were constructed to have identical $\Lambda_{1.4}$=193. The solid lines correspond to EoSs with $R_{1.4}=12$~km, and the dotted lines correspond to the $R_{1.4}=13$~km EoSs.}
\end{figure}

\subsubsection{Cold EoS in $\beta$-equilibrium}
\label{sec:coldEOS}
We model the zero-temperature, $\beta$-equilibrium EoS at densities above $0.5\ns$ with piecewise polytropes, as in Refs.~\cite{Ozel:2009da, Read:2008iy, Steiner:2010fz, Raithel:2016bux}. We use five polytropic segments, which are spaced uniformly in $\log{n}$ between 0.5 and 7.4~$\ns$.   In order to ensure a smooth matching between the low- and high-density EoSs, we fix the polytopic pressure at $0.5\ns$ to that of SFHo. The pressures at the remaining fiducial densities serve as free parameters, which we vary to construct EoSs with particular values for the slope of the symmetry energy, the neutron star radius, and the neutron star tidal deformability. We require that the maximum mass of each EoS is at least $2\Ms$, in order to satisfy observational constraints from massive pulsars \cite{Demorest:2010bx,Antoniadis:2013pzd,Fonseca:2016tux,NANOGrav:2019jur}. Additionally, we require that each EoS remains causal, and we set a lower limit on the pressures at the first two fiducial densities ($P(0.86\ns)>1.07$~MeV/fm$^{3}$ and $P(1.5\ns)>3.96$~MeV/fm$^{3}$ ), which correspond to pressures obtained from the Argonne AV8 two-body potential \cite{Gandolfi:2013baa}. This provides a lower limit to the low-density pressure, under the assumption that the three-nucleon interaction is always repulsive. Because the expansion of nuclear interactions to few-body potentials breaks down at higher densities, we only impose these constraints on our first two fiducial densities. Additionally, for one EoS in our sample ($R_{1.4}=10.8$~km, $L=$40~MeV; see below), we relax the lower limit at $P(1.5\ns)$ by 20\%, in order to explore a broader region of parameter space. For additional details on the choice of these EoS constraints, see \cite{Ozel:2015fia,Raithel:2016bux}.

Even though this parameterization is an agnostic way of describing the EoS, the polytropic pressure at $\ns$ still uniquely determines $L$, which we show as follows.  We start with the general expression for the energy per baryon of zero-temperature nuclear matter,
\begin{equation}
\label{eq:Enucl}
E_{\rm nucl}(n,Y_e, T=0) = E_0(n) + E_{\rm sym}(n)(1-2 Y_e)^2
\end{equation}
where $n$ is the baryon number density, $Y_e$ is the electron fraction, and $E_0(n)$ is the energy of symmetric nuclear matter.
The corresponding pressure is thus
\begin{multline}
\label{eq:Pnucl_full}
P(n, Y_e, T=0) = n^2 \left[ \frac{\partial E_0(n)}{\partial n}  \right] +\\
	n^2 \left[ \frac{\partial E_{\rm sym}(n)}{\partial n} \right]  (1-2 Y_e)^2.
\end{multline}
In this derivation, we neglect the contribution of electrons, which add a $\lesssim10$\% correction into the determination of $L$ below. At the nuclear saturation density, $\partial E_0(n)/\partial n$ is zero by definition, and so the first term in eq.~(\ref{eq:Pnucl_full}) vanishes. In order to simplify the second term, we use the fact that, for matter in $\beta$-equilibrium, the electron fraction is completely determined by the symmetry energy, i.e., $Y_{e,\beta} = Y_{e,\beta}(n, S, L)$, to leading-order in the symmetry energy expansion \cite[see, e.g., Appendix A of][]{Raithel:2019gws}. Following \cite{Raithel:2019ejc}, we approximate the $\beta$-equilibrium neutron excess as
\begin{equation}
\label{eq:Yp_approx}
\left(1-2Y_{e,\beta}\right)^2 = a + b u + \mathcal{O}(u^2)
\end{equation}
where $u\equiv (n/\ns)-1$, $a=a(S)$, $b=b(S,L)$, and we have suppressed the dependencies of $Y_{e,\beta}$ on the density and symmetry energy parameters for clarity. With this approximation, we thus have $(1-2 Y_{e,\beta})^2|_{\ns} \approx a$. Finally, we can further simplify eq.~(\ref{eq:Pnucl_full}) by substituting in $L\equiv 3\ns \left( \partial E_{\rm sym}/\partial{n}\right)|_{\ns}$, which follows from the definition in eq.~(\ref{eq:Esym}). Combining these results, the $\beta$-equilibrium pressure at $\ns$ is given by
\begin{equation}
\label{eq:PofL}
P(\ns, Y_{e, \beta}, T=0) = \frac{a\ns L}{3},
\end{equation}
or
\begin{equation}
\label{eq:LofP}
L = \frac{ 3 P(\ns, Y_{e, \beta}, T=0) }{a \ns}.
\end{equation}
For a similar derivation, see \cite{Raithel:2019ejc}. 
For the EoSs constructed in this paper, we fix $S$=32~MeV, in order to be consistent with recent theoretical and experimental constraints \cite{Li:2021thg}. Accordingly, $a(S)=0.833$ \cite{Raithel:2019ejc}.

Using eq.~(\ref{eq:PofL}), we vary the pressures in our piecewise polytropic model to fix $L$ to either 40, 100, or 120~MeV. These values were chosen in order to span the range of constraints from astrophysics and from the recent PREX-II measurement \cite{Li:2021thg}. Fixing $L$ effectively sets the pressures at the first two fiducial densities, which bracket $\ns$. We then vary the remaining pressures in the polytropic model to construct sets of EoSs that match in either $R_{1.4}$ or $\Lambda_{1.4}$. We show the resulting sample of seven EoSs in Fig.~\ref{fig:MR}, along with the corresponding mass-radius and tidal deformability curves.

\begin{table*}[!ht]
\centering
\begin{tabularx}{0.8\textwidth}{@{\extracolsep{\fill}}ccccccccccc}
\hline \hline
Approx. radius &  $M_{\rm tot} [\Ms]$   & $q$ & $L$~[MeV]  & $R_{1.4}$~[km]& $R_{1.8}$~[km]&  $\Lambda_{1.4}$ & $\widetilde{\Lambda}$  & $M_{\rm max}~[\Ms]$ \\
\hline 
			                	&&   &40  & 10.8  & 10.8	& 193  & 237 &     2.14	  \\  	
 	R $\simeq$ 11 & 2.72& 0.85 &     100  &  11.2 &  11.1 & 193 & 241   & 2.14   \\ \hline

	  			&&  	&    40 & 12.0  &12.3  & 425 &  517, 537 & 2.47	 \\ 	
 	R = 12  &  2.72, 2.71 & 0.85, 1  &  100 &  12.0  & 11.9      & 311   & 394, 395	 & 2.25 \\
				        && & 120 &  12.0  & 11.6  & 287 &   364, 372 	& 2.20  \\ \hline
				
	 			&& &  100 & 13.0  	& 12.5 & 557  &  699  & 2.14  \\	
 	R = 13	& 2.72& 0.85 &	  120 &   13.0 & 12.6      & 522  & 662   & 2.23 \\ \hline
\hline
\end{tabularx}
\caption{Summary of equation of state (EoS) and binary configurations explored in this work.\\
Here, $M_{\rm tot}$ is the total gravitational mass at infinite separation of the binary,
$q$ its mass ratio and $\tilde{\Lambda}$ its effective tidal deformability of the binary, defined as in eq. (5) of \cite{Favata:2013rwa} .
The EoS parameters are given by the slope, $L$, of the nuclear symmetry energy; the radii, $R_{1.4}$ and $R_{1.8}$, of a $1.4$ and $1.8,M_\odot$ neutron star (NS), respectively; the tidal deformability of $1.4\, M_\odot$ NS, $\Lambda_{1.4}$; and the maximum mass of a nonrotating NS, $M_{\rm max}$ .}
  \label{table:EoSs}
\end{table*}

Our final sample of EoSs contains three subsets which are designed for systematic comparison. In the first subset, we construct three EoSs that all predict $R_{1.4}=12$~km, but that span the full range of $L=40,$ 100, and 120~MeV. Because we span all three values of $L$ for this sample, these models will be the main focus of this paper. We additionally construct a set of two stiffer EoSs that predict a larger radius of $R_{1.4}=13$~km, for $L=100$ and 120~MeV. We find that it is not possible to construct a model with $R_{1.4}=13$~km and $L=40$~MeV without violating causality, within the particular polytropic framework used in this work. As a result, the $R_{1.4}=13$~km EoSs span just the larger values of $L=100$ and 120~MeV. Finally, we construct a set of softer EoSs, which were designed to match exactly in $\Lambda_{1.4}$, rather than in their radii. This allows us to study whether varying the tidal deformability, radius, or $L$ has a larger impact on the post-merger properties.  For the softer set of EoSs, we focus on $L=40$ and 100~MeV, with $\Lambda_{1.4}=193$ for both cases, and $R_{1.4}=10.8$ and 11.2~km, respectively. The complete sample is shown in Fig.~\ref{fig:MR}, and their characteristic properties are summarized in Table~\ref{table:EoSs}.

\subsubsection{Extrapolation to finite temperatures and arbitrary electron fraction}
The piecewise polytropic framework is used to characterize zero-temperature, $\beta$-equilibrium matter at densities above $0.5\ns$. However, in a neutron star merger, the post-merger temperatures can reach several tens of MeV and the electron fraction can also deviate from the initial $\beta$-equilibrium composition \cite[e.g.,][]{Oechslin:2006uk,Sekiguchi:2011zd,Bernuzzi:2015opx,Perego:2019adq}. In this section, we describe the key features of our extrapolation of the piecewise polytropes to finite temperatures and arbitrary electron fraction, which follows the framework of Ref.~\cite{Raithel:2019gws}. For complete details on how the pressure and energy are calculated at fixed $n$, $T$, and $Y_e$, see Boxes~I and II of that work.

We extrapolate to finite-temperatures using the $M^*$-model, which provides an approximation of the thermal pressure, including the leading-order effects of degeneracy at high densities \cite{Raithel:2019gws}. For all EoSs constructed in this paper, we use an identical set of $M^*$-parameters, $n_0$=0.12~fm$^{-3}$ and $\alpha=0.8$, which are consistent with the values inferred from a sample of nine of finite-temperature EoS tables \cite{Raithel:2019gws}. 

The extrapolation from $\beta$-equilibrium to arbitrary $Y_e$ utilizes the leading order expansion coefficients of the symmetry energy, $S$ and $L$, as well as an additional parameter $\gamma$, which characterizes the density dependence of interactions between the particles. For all EOSs in our sample, we fix $S$=32~MeV, as described in Sec.~\ref{sec:coldEOS}. The slope $L$ is set according to eq.~(\ref{eq:LofP}), and we choose $\gamma=0.6$, which is consistent with the range of values inferred from tabulated EoSs \cite{Raithel:2019gws}. 

We note that, for $n\lesssim0.5\ns$, the nuclear symmetry energy framework breaks down, due to the formation of bound nuclei. This complicates the extrapolation from  $\beta$-equilibrium to arbitrary $Y_e$, which is grounded in the symmetry energy formalism. In order to avoid these issues, we switch to the tabulated EoS SFHo at densities below $0.5\ns$. The matching of this low-density EoS to our high-density, finite-temperature models is performed following \cite{Schneider:2017tfi}, with a transition window from $n=6.3\times10^{-5}$ to 0.08~fm$^{-3}$ (see also Ref. \cite{Most:2018eaw})).
Additionally, across this transition window, we switch from the complete model for $E_{\rm sym}(n)$ to a decaying power-law function, with parameters that are chosen to ensure that the extrapolation to arbitrary $Y_e$ remains realistic across the window.  For additional details, see Appendix~\ref{sec:Esym_lown}. 

The nuclear symmetry energy expansion formalism is also expected to break down at very high densities of a few times $\ns$, where additional degrees of freedom may become important. We do not account for non-nucleonic degrees of freedom in this paper, and leave the exploration of such effects to future work.

Finally, we note that in addition to the pressure, energy, and sound speed (which are all calculated following Ref.~\cite{Raithel:2019gws}), the numerical evolution also requires input for the chemical potentials, which are used to model the neutrino transport (see Sec.~\ref{sec:numerics}). We describe the calculation of the chemical potentials in Appendix~\ref{sec:mu}.

\begin{figure*}
    \centering
    \includegraphics[width=\textwidth]{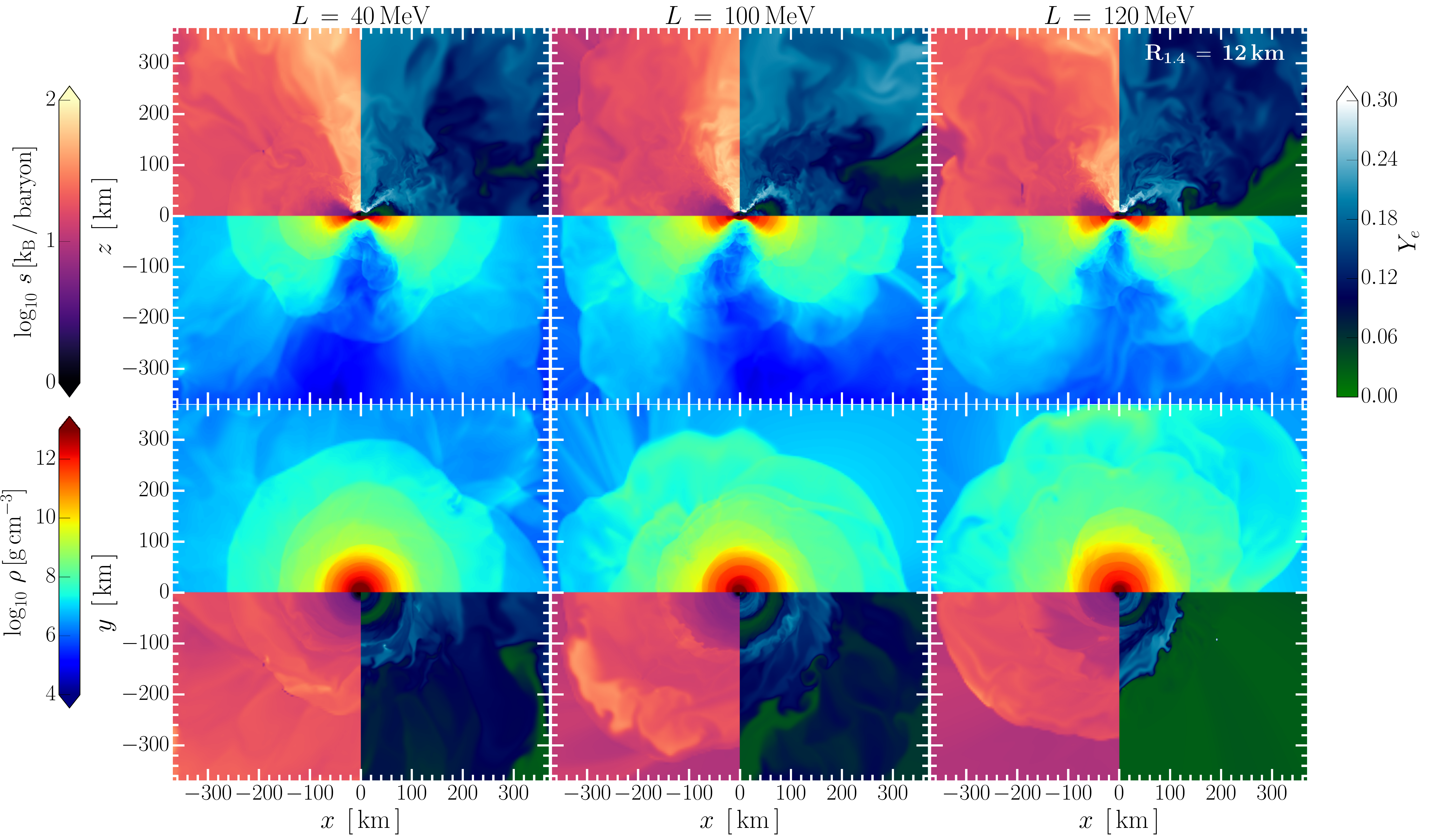}
    \caption{Two-dimensional spatial distributions of the rest-mass density $\rho$, temperature $T$, electron fraction $Y_e$, and entropy $s$ per baryon. Shown are meridional (top) and equatorial (bottom) views for three different values of the slope $L$ of the nuclear symmetry energy. The results shown are for the $R_{1.4}\,=\,12\,\rm km$ models with mass ratio $q=0.85$ at  time $t=25\, \rm ms$ after merger.}
    \label{fig:results_overview}
\end{figure*}

\subsection{Numerical Setup}
\label{sec:numerics}
In this work, we simulate the final orbits, merger, and post-merger phase of a
binary neutron star coalescence. This requires us to model both the evolution of the
fluid as well as the self-consistently coupled dynamical evolution of the
space-time.
For the latter, we solve the equations of general relativity using the Z4c
\cite{Hilditch:2012fp,Bernuzzi:2009ex} formulation, which is based on the
Z4 formulation \cite{Bona:2003fj}, that allows for a dynamical damping of constraint
violations to the Einstein field equations \cite{Gundlach:2005eh}.
Following \cite{Weyhausen:2011cg}, we choose damping parameters
$\kappa_1=0.02$ and $\kappa_2=0$. The gauge conditions adopt moving
puncture gauges, i.e. 1+log slicing with Gamma-driver conditions
\citep{Alcubierre:2002kk}, with damping parameter $\eta=0.5$.
We further find it beneficial to add an inverse radial fall-off to the
damping parameters outside of a sphere of $r=500\, \rm km$, to preserve
numerical stability \cite{Schnetter:2010cz}. To damp high-frequency noise
in the gravitational field sector, we add Kreiss-Oliger dissipation
\cite{Babiuc:2007vr}.\\

On the dynamically evolved background, we solve the equations of ideal
general-relativistic (magneto-)hydrodynamics (GRMHD)
\cite{Duez:2005sf,Shibata:2005gp}, in the limit of vanishing magnetic field
strength. Weak interactions are included following the leakage prescription
outlined in \cite{Ruffert:1995fs,Rosswog:2003rv}, which accounts for local
energy losses and composition changes due to neutrino interactions.\\

We solve these equations using the Frankfurt-/IllinoisGRMHD code (\texttt{FIL})
\cite{Most:2019kfe,Most:2018eaw}, which is derived from the publicly
available \texttt{IllinoisGRMHD} code ( \texttt{ILGRMHD} ) \cite{Etienne:2015cea}.
To solve the Einstein equations, \texttt{FIL} provides a fourth-order accurate numerical implementation
of the Z4c system using the methods outlined in
\cite{Zlochower:2005bj}.\\
Different from \texttt{ILGRMHD}, \texttt{FIL} utilizes a fourth-order
accurate conservative finite-difference algorithm based on the ECHO scheme 
 to solve the GRMHD equations \cite{DelZanna:2007pk}.
 Crucially for this work, \texttt{FIL} provides its
own microphysics infrastructure, which can handle fully tabulated EoSs.
The codes makes use of the publicly available \texttt{Einstein Toolkit}
infrastructure \citep{Loffler:2011ay} and specifically the \texttt{Carpet}
moving boxes refinement code \cite{Schnetter:2003rb}.
Specifically, we set up our simulation domain to extend to an outer
boundary of $3022\, \rm km$ in each direction and to consist of 8
refinement levels of doubling resolution, where the finest one covering the
stars has a resolution of $262\,\rm m$.
For computational efficiency, we employ reflection symmetry across the
vertical $z$-direction.

\subsection{Initial conditions}
\label{sec:id}
We model the initial irrotational neutron star binaries in quasi-circular equilibrium 
\cite{Gourgoulhon:2000nn} using the \texttt{LORENE} code.\footnote{https://lorene.obspm.fr}
The two neutron stars are placed at an initial separation of $45\, \rm km$ and are
constructed for each of the EoSs outlined in Sec. \ref{sec:EOS}.
The binary parameters are modelled after the GW170817 event
\cite{TheLIGOScientific:2017qsa,Abbott:2018wiz}. In particular we adopt two
mass ratios $q=\left[ 0.85; 1.0 \right]$, where the latter is only used
with $R_{1.4}=12\,\rm km$ EoSs.
This fixes the total mass $M$ of the system, via the chirp mass $\mathcal{M}= M q^{3/5}/\left(1+q  \right)^{6/5}
= 1.186\, M_\odot$ \citep{Abbott:2018wiz}. 

\begin{figure*}[!t]
\centering
\includegraphics[width=\textwidth]{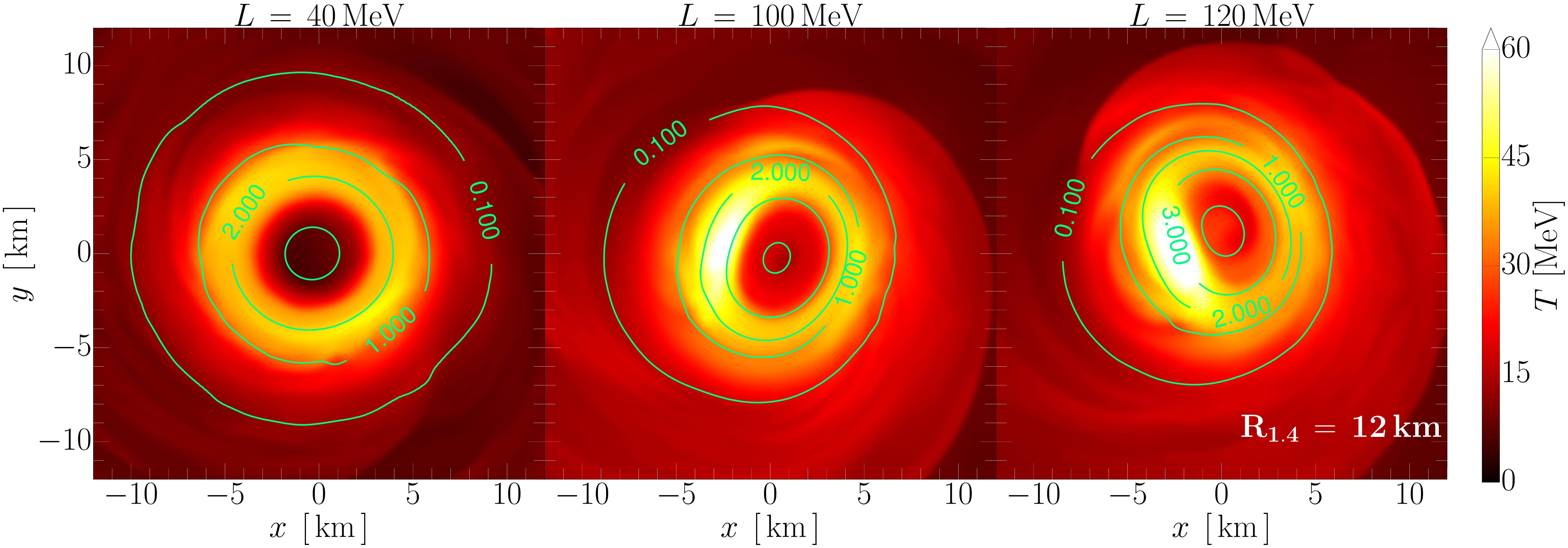}
\caption{\label{fig:2D_temp} Temperature $T$ in the equatorial plane at $t\simeq$ $20\, \rm ms$ after merger for unequal mass $(q=0.85)$ mergers with EoSs having a characteristic radius of $R_{1.4}\,=\, 12 \rm km$. The green lines indicate contours of constant rest-mass density, with values labelled with respect to the nuclear saturation density. The different panels show results for varying slope parameter $L$ from 40 to 120~MeV.}
\end{figure*}

\section{Results}
\label{sec:results}
We now turn to the results of our simulations, with an emphasis
on the unequal mass case $(q=0.85)$, which most closely matches the masses inferred from GW170817. 
Starting from a description of the general dynamics and remnant properties,
we will present a discussion of the mass ejecta and of the gravitational
wave signals expected for various choices of the slope of the symmetry
energy $L$.\\

Since the main goal of this paper is to determine how $L$ affects the early post-merger phase, we focus our attention on 
the dynamics in the first few tens of milliseconds after merger.
During this time, high temperatures will be reached during merger
\cite{Oechslin:2006uk,Sekiguchi:2011zd,Bernuzzi:2015opx,Perego:2019adq},
post-merger bounces can drive an early mass ejection \cite{Nedora:2020qtd}, and the rotating remnant will emit significant GWs. At the end of this process, the former two neutron star cores
will have fused into a single core, with continued, diminishing gravitational wave
emission leading to an axisymmetrization of the remnant. 
This newly formed massive neutron star will be hot and
rapidly, differentially rotating \cite{Hanauske:2016gia}, unless strong shear viscosity is present \cite{Radice:2017zta,Shibata:2017xht}. This remnant will set the stage for
long-term mass ejection and neutrino emission. 

For all EoSs considered in this paper except one, we find that the remnants survive until the end of our simulations ($t\simeq 25$~ms post-merger). The sole exception is the EoS with $L\,=\,40\, \rm MeV$ and $R_{1.4}\,=\, 10.8\, \rm km$, which collapses after $15\, \rm ms$. When comparing to this EoS, we will accordingly limit our comparisons to the first $15\, \rm ms$ post-merger. For all other EoSs, we will present results from the end of our simulations (about $25\,\rm ms$ post merger), unless otherwise specified.

In order to provide a first indication of how the post-merger remnant and  early mass ejection depend on the slope of the symmetry energy,
Fig. \ref{fig:results_overview} shows equatorial and meridional cuts
of the rest-mass density, $\rho$, specific entropy, $s$, and electron fraction,
$Y_e$, about $25\, \rm ms$ after merger. 
To aid the comparison, we focus here on models
with radii $R_{1.4}\,=\, 12\, \rm km$. Starting out with the rest-mass density,
in the equatorial plane (bottom row of Fig. \ref{fig:results_overview} ) 
we can see clear differences in the total amount of mass ejection.
The $L\,=\, 120\, \rm MeV$ case shows extended shock fronts with densities
$\rho < 10^{10}\, \rm g\, cm^{-3}$, that are reduced in size for the
$L\,=\,40\, \rm MeV$ case. Additionally, the $L\,=\,40\, \rm MeV$ profiles are
much more axisymmetric than for higher values of $L$. 

Looking at the
electron fractions $Y_e$, we find that the disk and most of the ejecta
are very neutron rich, i.e. $Y_e < 0.02$ (green regions) for the large $L$ EoSs, whereas the proton fraction is slightly enhanced, $Y_e\simeq 0.1$, for low $L$. The electron fractions in the polar region are overall comparable between the EoSs,
but are slightly more proton rich in the $L\,=\,40\, \rm MeV$, which is related to much higher shock heating, as evidenced by the enhanced specific entropies $s$ for this EoS.
Specifically, we also find differences in the specific entropies reached for each of our EoSs, as shown in the pink panels of Fig.~\ref{fig:results_overview}. We find that the merger with $L\,=\, 100\rm MeV$ reaches the highest specific entropy, in both the equatorial plane and along the polar axis. However, for all values of $L$, we find that $s$ can exceed 10~$\rm{k_B/baryon}$, in the low-density outflows.
Since $s$ is a proxy for the
amount of shock heating taking place, we are led to conclude that shock
heating will be important for all values of $L$, for the $R_{1.4}\,=\,12\rm
km$ models in our sample. We discuss these properties of the ejecta in further detail below.

\begin{figure*}[!ht]
\centering
\includegraphics[width=\textwidth]{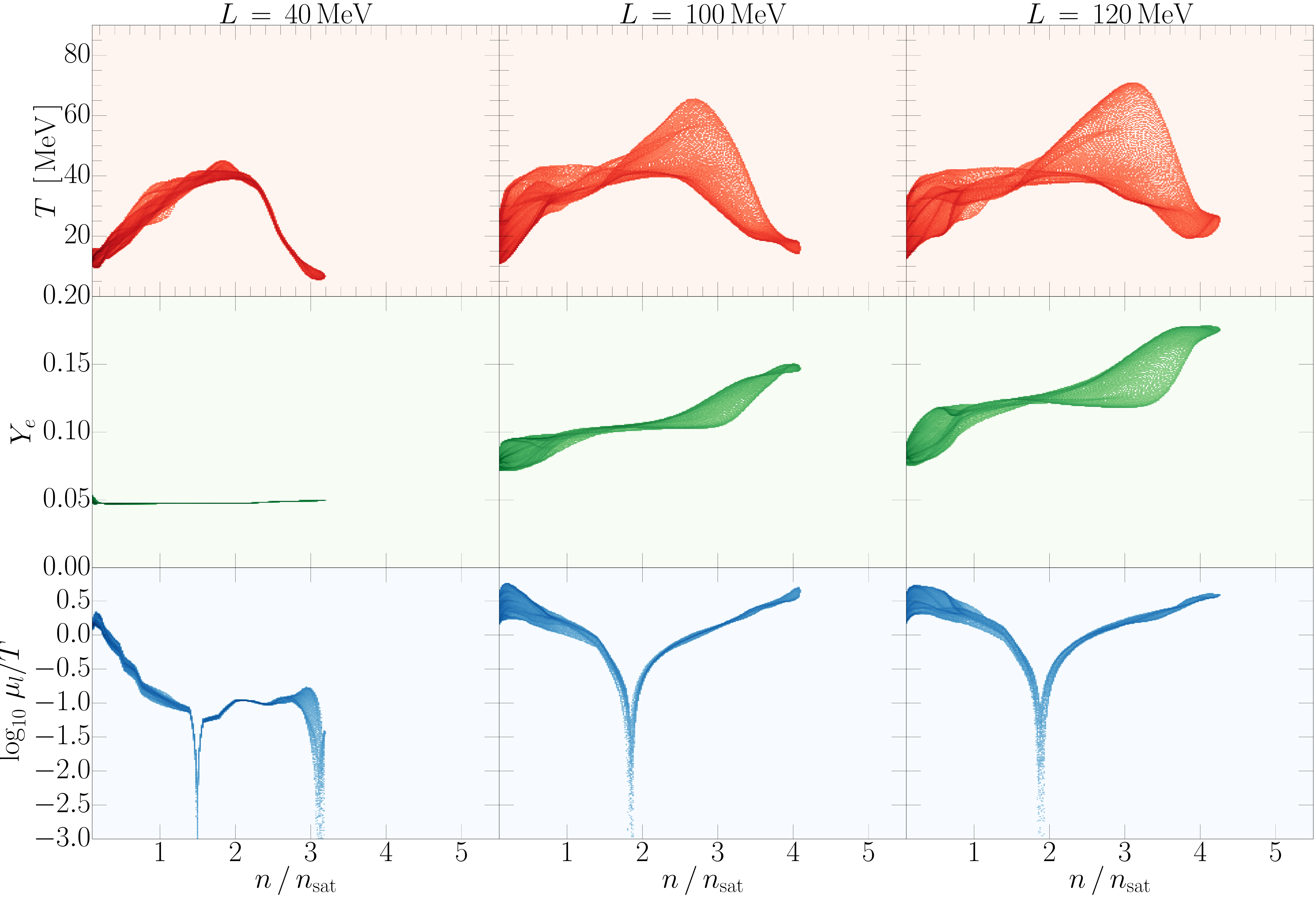}
\caption{\label{fig:nT} Temperatures $T$, electron fractions $Y_e$ and lepton chemical potential $\mu_l$ probed at different densities $n$ in the massive neutron star remnant. The densities are stated relative to saturation density $n_{\rm sat}$. The models are the same as shown in Fig. \ref{fig:2D_temp}.}
\end{figure*}

\subsection{Remnant properties}
\label{sec:remnant}
We now turn to the general properties of the hypermassive neutron star remnant, 
and how it is impacted by differences in the slope $L$ of the nuclear symmetry energy.
Since we have three different slope values available for the $R_{1.4}\,=\,12\rm km$ case (see Sec. \ref{sec:EOS}), we will mainly focus on these models in this section.\\

We begin by discussing the overall structure and thermodynamic conditions present in the hypermassive neutron star.
Figure \ref{fig:2D_temp} shows the temperature and density distribution in the equatorial plane, at $\simeq 20\, \rm ms$ after merger. We can see that there are several differences between the models.
Starting for the $L\,=\, 40\, \rm MeV$ model (left panel), we see that
after merger temperatures of about $40\,\rm MeV$  are reached in a hot ring
with densities between $1-2\, n_{\rm sat}$.
Surprisingly, despite starting from an initially asymmetric merger, the system has quickly circularized. This behavior will be more closely examined in Sec. \ref{sec:GW} in the context of the associated gravitational wave emission and the decay of the $m=1$ component.
For the $L=\,100$ and $120\, \rm MeV$ cases, we instead find that the hot ring is highly asymmetric, with temperatures $>60\, \rm MeV$ being reached in parts of the hot ring. Similar to what has been found by varying the finite-temperature part of the EoS \cite{Raithel:2021hye}, we find that changes in $L$ lead to different temperatures in the colder center of the star. This suggests that, at least within some part of the parameter space, finite temperature effects might be degenerate with changes to the cold EoS, in determining the thermal profile of the remnant.  Such temperature differences might be crucial when determining the microphysical conditions necessary for (bulk-) viscous effects to become important \cite{Most:2021zvc}, and may also influence the local neutrino emissivity of the remnant and, as a result, the cooling and eventual neutrino irradiation of the disk. 

Overall, we find that the maximum temperatures reached in the merger correlate strongly with the initial neutron star radius, as summarized in Table~\ref{table:ejecta}. Mergers with $R_{1.4}\,=\,11\,\rm km$ reach temperatures above $120\,\rm MeV$ at merger, whereas those with $R_{1.4}\,=\,13\,\rm km$ only reach temperatures $\lesssim 100\,\rm MeV$. This is consistent with previous findings that more compact neutron stars collide with higher impact velocities \cite{Bauswein:2013yna}, and thus would be expected to experience stronger shock heating. At late times within the massive neutron star remnant, however, the temperatures are less strongly correlated with $R_{1.4}$, as can be seen in Fig.~\ref{fig:2D_temp}, where for all EoSs with $R_{1.4}=12$~km, the maximum temperatures vary between $40 \lesssim T^{\rm max}_{\rm final} \lesssim 70$ MeV.  Instead, we find that the maximum temperature of the late-time remnant correlates weakly with the radii at high masses, i.e. for $M\,\ge\, 1.8\, M_\odot$. That is, EoSs with small $R_{1.8}$ overall reach temperatures above $60\, \rm MeV$, whereas EoSs with larger ($R_{1.8}>12\, \rm km$), only reach temperatures $< 50\, \rm MeV$. 
Therefore, it seems likely that the late-time remnant temperatures are, at least in part, governed by the high density part of the cold EoS. We list the values of $R_{1.8}$ in Table~\ref{table:EoSs} for reference, while the maximum merger and late-time temperatures are summarized in Table~\ref{table:ejecta}.

We continue our description of the remnant temperature by studying the distribution of temperatures in terms of the densities at which they occur, again focusing on the $R_{1.4}=12$~km models. Previous studies have considered these conditions either in the general thermodynamics of the merger \cite{Perego:2019adq,Hanauske:2019qgs} or the appearance of a hot quark-matter phase \cite{Most:2019onn,Prakash:2021wpz}. In the top row of Fig. \ref{fig:nT}, we now directly compare the thermodynamic conditions present at a given density. We can see that due to differences in the cold EoS, and hence in $L$, different densities are reached. In particular, the $L=40\, \rm MeV$ merger probes lower densities of around $3\, \rm n_{\rm sat}$, while higher values of $L$ lead to densities beyond $4\, \rm n_{\rm sat}$ being probed in the post merger. This is a direct result of the overall stiffness of these EoSs. Most strikingly, in the case of the $L\,=\,40\,\rm MeV$ EoS, the temperature distribution is very narrow, and  follows an almost univariate profile with the density. On the other hand, the temperatures probed in the $L\,\geq\, 100\, \rm MeV$ cases are more broadly distributed, with temperatures of up to $70\, \rm MeV$ being reached. Also in these cases, a simple mapping between $n$ and $T$ is no longer possible, as the range in temperatures can be quite broad for a given density, e.g. $30\, \rm MeV$ differences for matter at $n\simeq 3\, n_{\rm sat}$.\\

At the same time, we can also compare how the electron fraction of the system varies, which would be a proxy for out-of-weak-equilibrium effects. While initially the electron fraction in the inspiral will be fixed at cold $\beta$-equilibrium, at high temperatures the conditions for $\beta$-equilibrium are modified. 
Indeed, we can see that for the coldest case, which corresponds to $L\,=\,40\,\rm MeV$, the electron fraction is almost constant with a very narrow distribution, with $Y_e \simeq 0.05$. For higher values of $L$, $Y_e$ increases to higher densities, as expected from cold $\beta$-equilibrium, and also attains a considerable spread, as a result of the large spread in $T$ for these EoSs. Nevertheless, even for the highest values of $L$ considered here, the electron fraction remains low, $Y_e<0.2$.

To reinforce these observations, Fig. \ref{fig:2D_temp} also shows the lepton chemical potential $\mu_l$, which vanishes in $\beta$-equilibrium.\footnote{We note that the sign change in $\mu_l$ indicates a relative increase or decrease of the electron fraction $Y_e$ relative to its $\beta$-equilibrium value.} From this, we can see that indeed out-of-equilibrium effects are likely small for the $L\,=\,40\,\rm MeV$ case, but may be significant for matter above $2\, n_{\rm sat}$ for larger L. This could have implications on the long-term thermal evolution and neutrino cooling of the remnant \cite{Fujibayashi:2017xsz,Fujibayashi:2017puw}.

\subsection{Mass ejection}
\label{sec:ejecta}
In this section we focus on the mass dynamically ejected during
the merger process, see e.g. \cite{Sekiguchi:2016bjd,Lehner:2016lxy,Bovard:2017mvn,Radice:2018pdn}. Although this is in most cases only a small contribution to
the overall amount of ejecta from the system \cite{Abbott:2017wuw}, the exact details typically
depend on the EoS and the mass ratios used in the simulation
\cite{Sekiguchi:2016bjd}. Additionally, the recent observation of an X-ray rebrightening
\cite{Hajela:2021faz,Balasubramanian:2021kny}, potentially associated with the presence of relativistic fast
ejecta \cite{Hotokezaka:2018gmo,Nedora:2021eoj}, has resulted in a recent interest into the properties of
dynamical mass ejection.

\begin{figure*}
\begin{center}
  \centering
  \includegraphics[height=3cm, trim=0 0 60 0, clip]{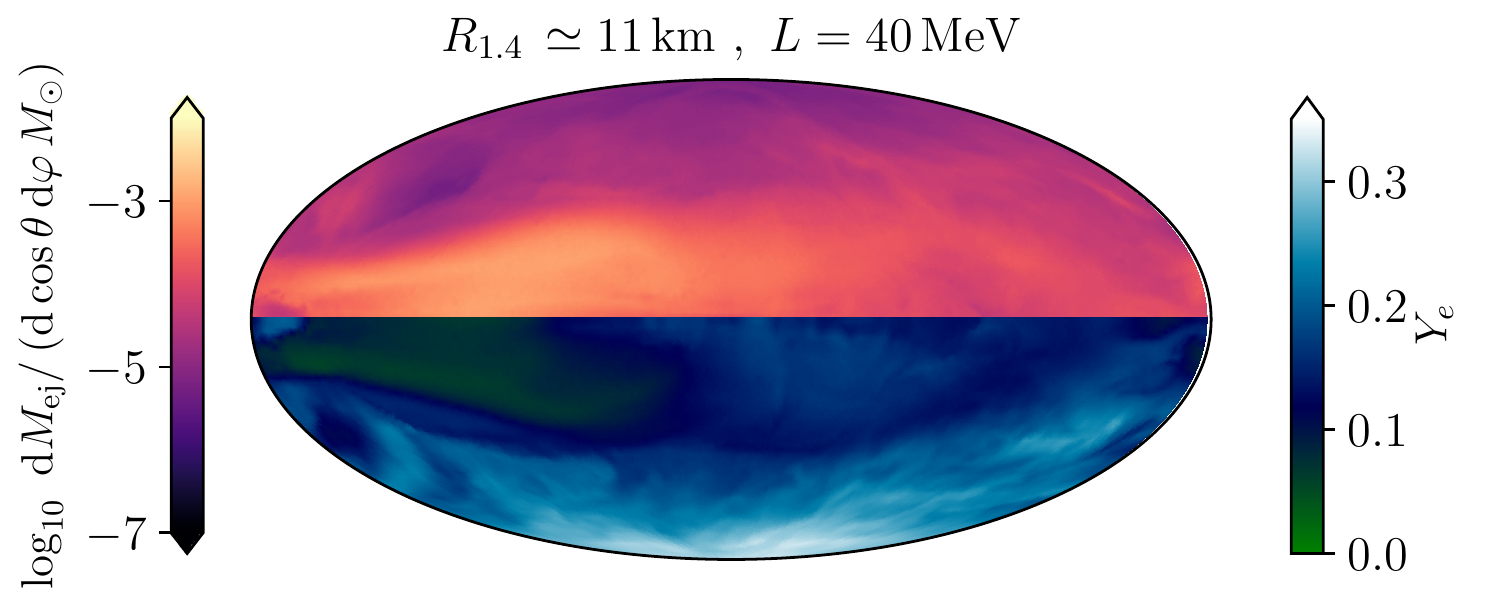}
  \includegraphics[height=3cm, trim=60 0 60 0, clip]{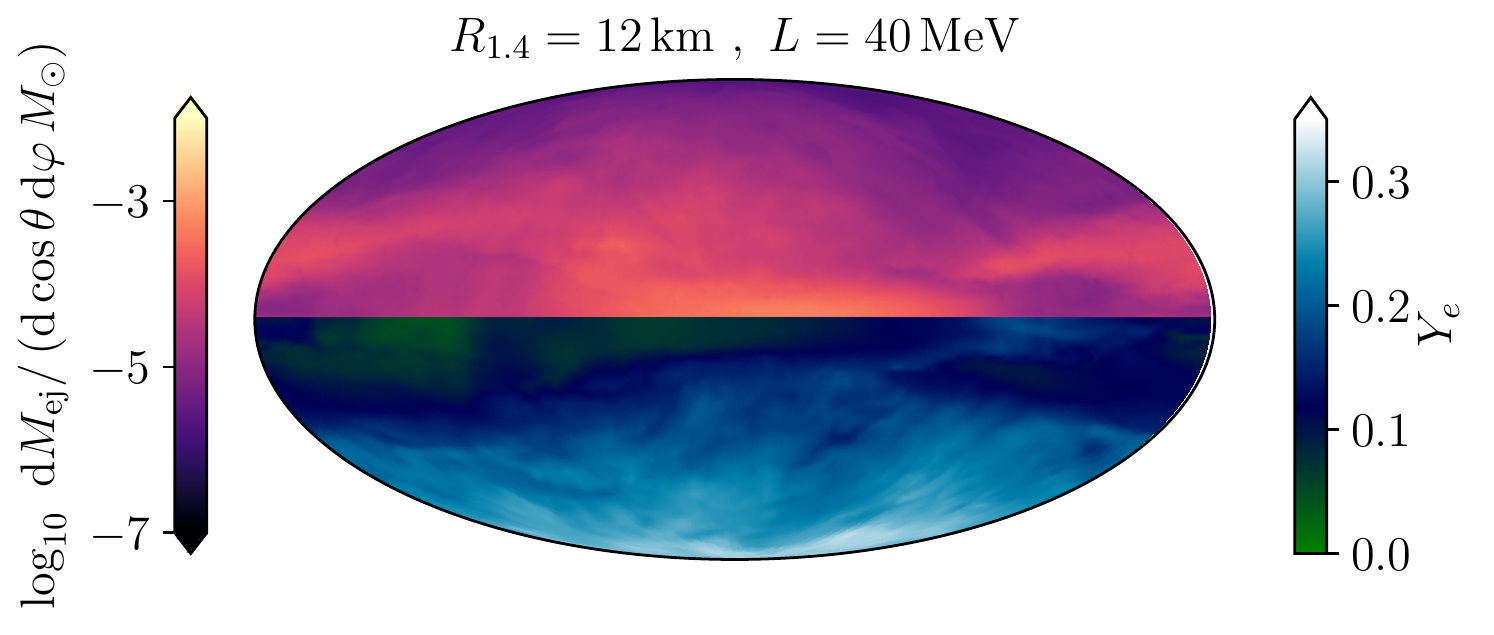}
  \includegraphics[height=3cm, trim=60 0 0 0, clip]{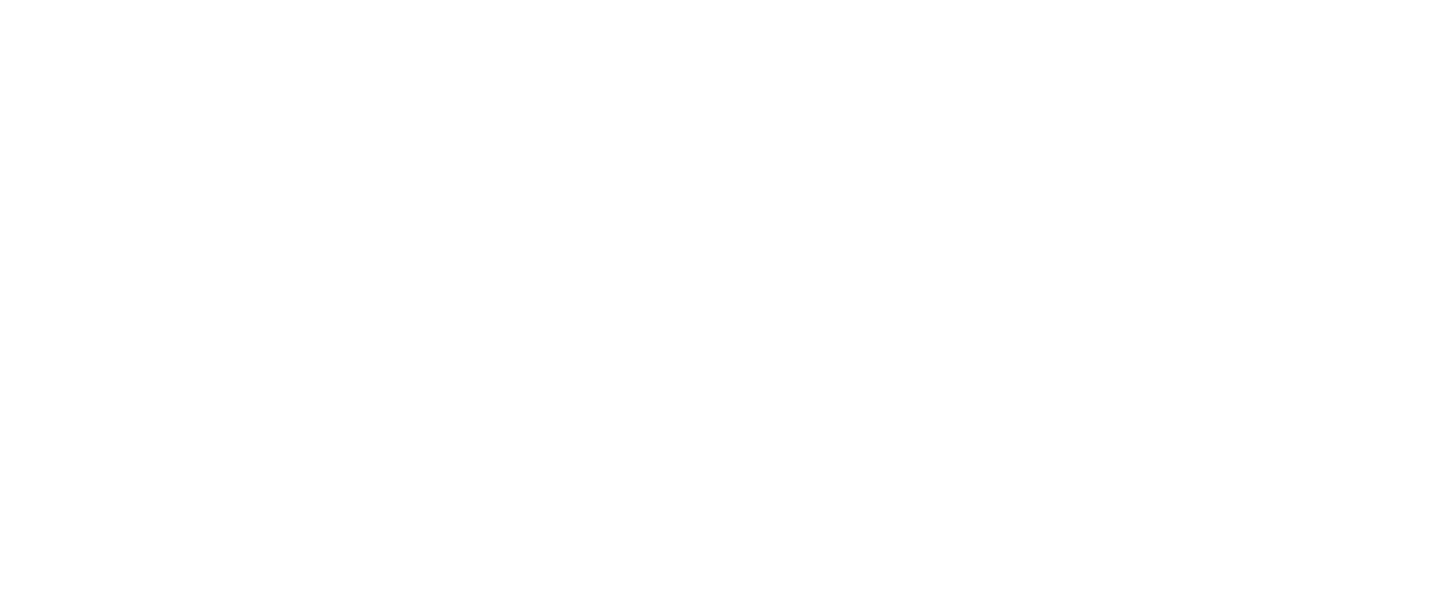}
  \\
  \includegraphics[height=3cm, trim=0 0 60 0, clip]{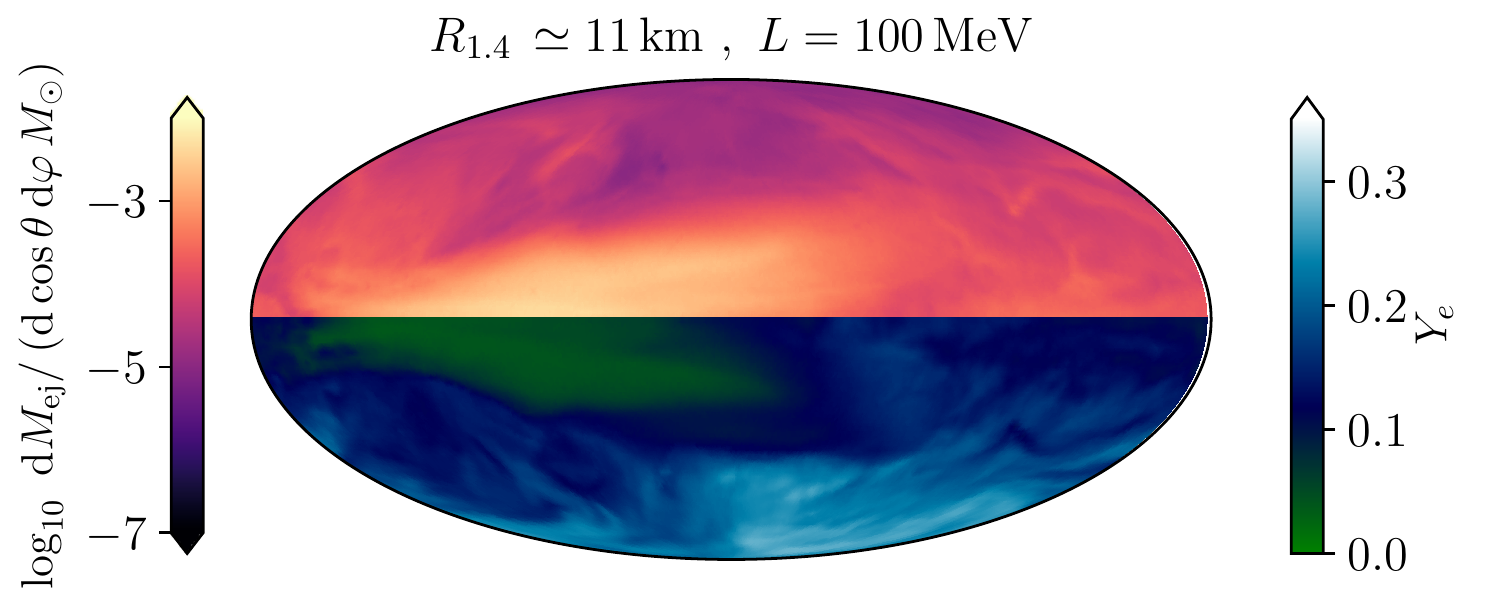}
  \includegraphics[height=3cm, trim=60 0 60 0, clip]{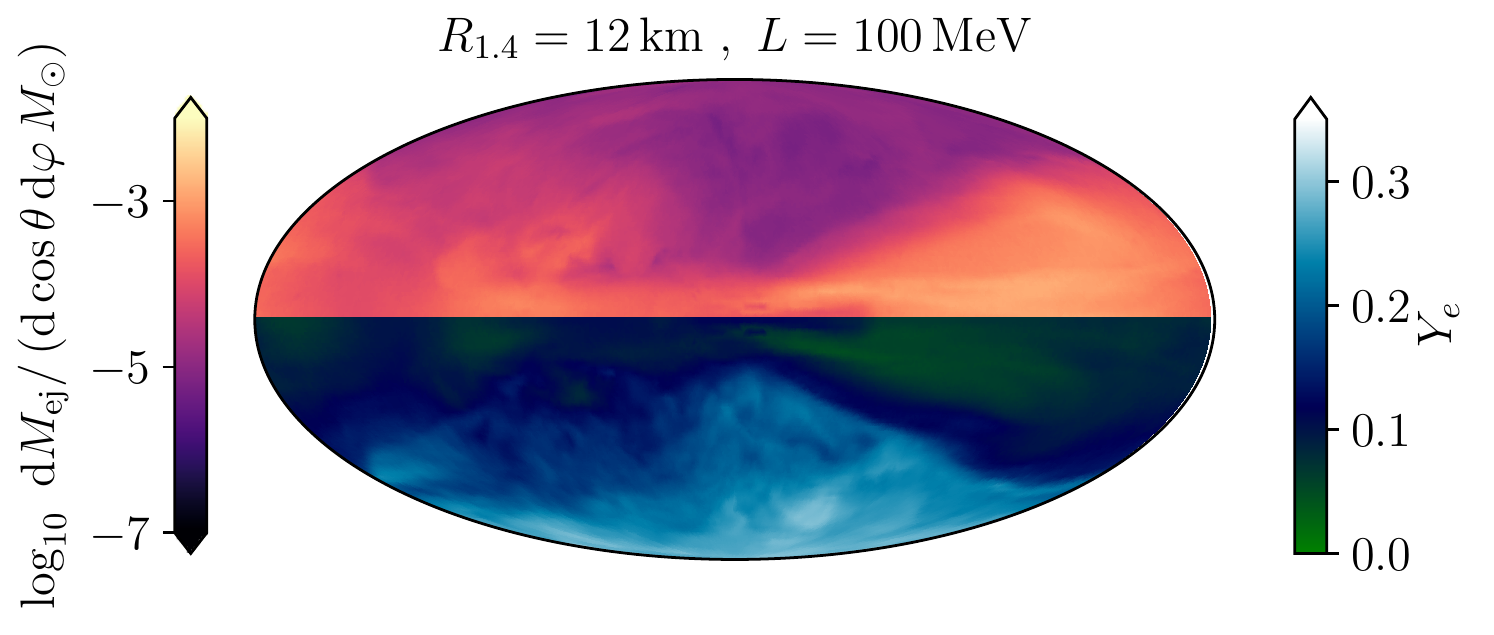}
  \includegraphics[height=3cm, trim=60 0 0 0, clip]{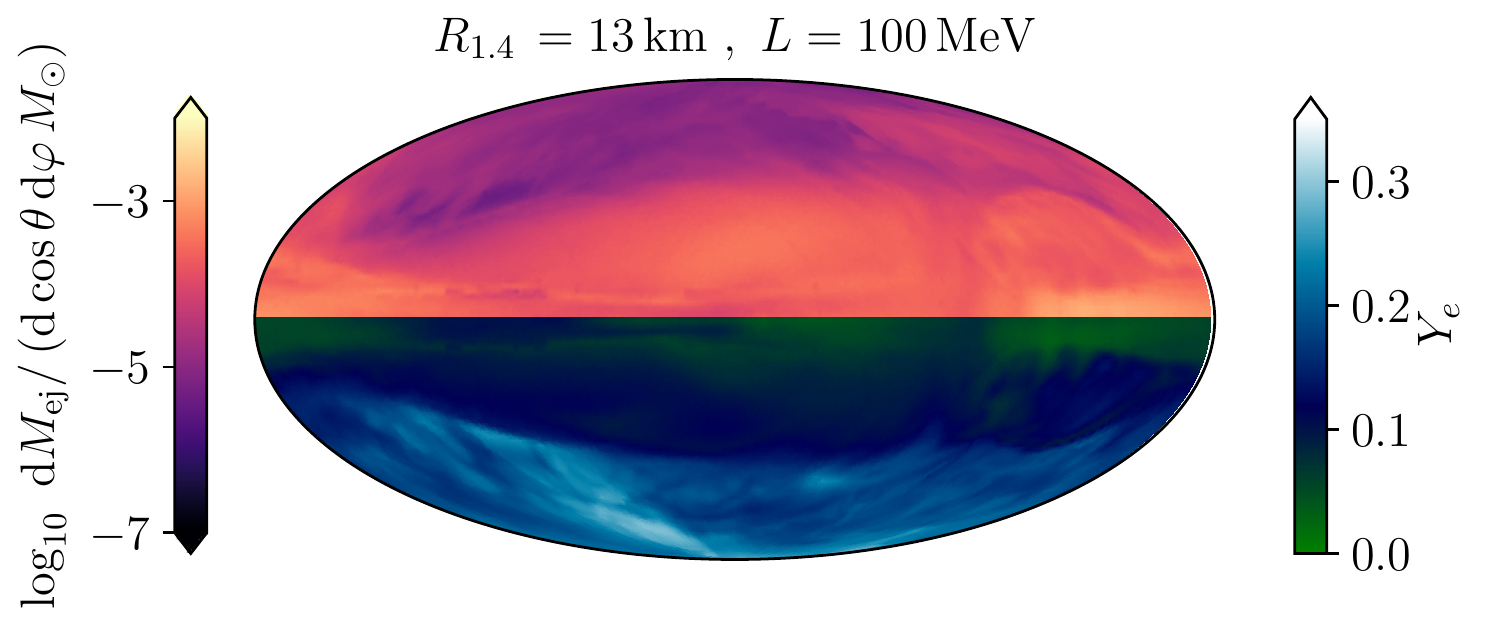}
  \\
 \includegraphics[height=3cm, trim=0 0 60 0, clip]{figures/dummy.pdf}
  \includegraphics[height=3cm, trim=60 0 60 0, clip]{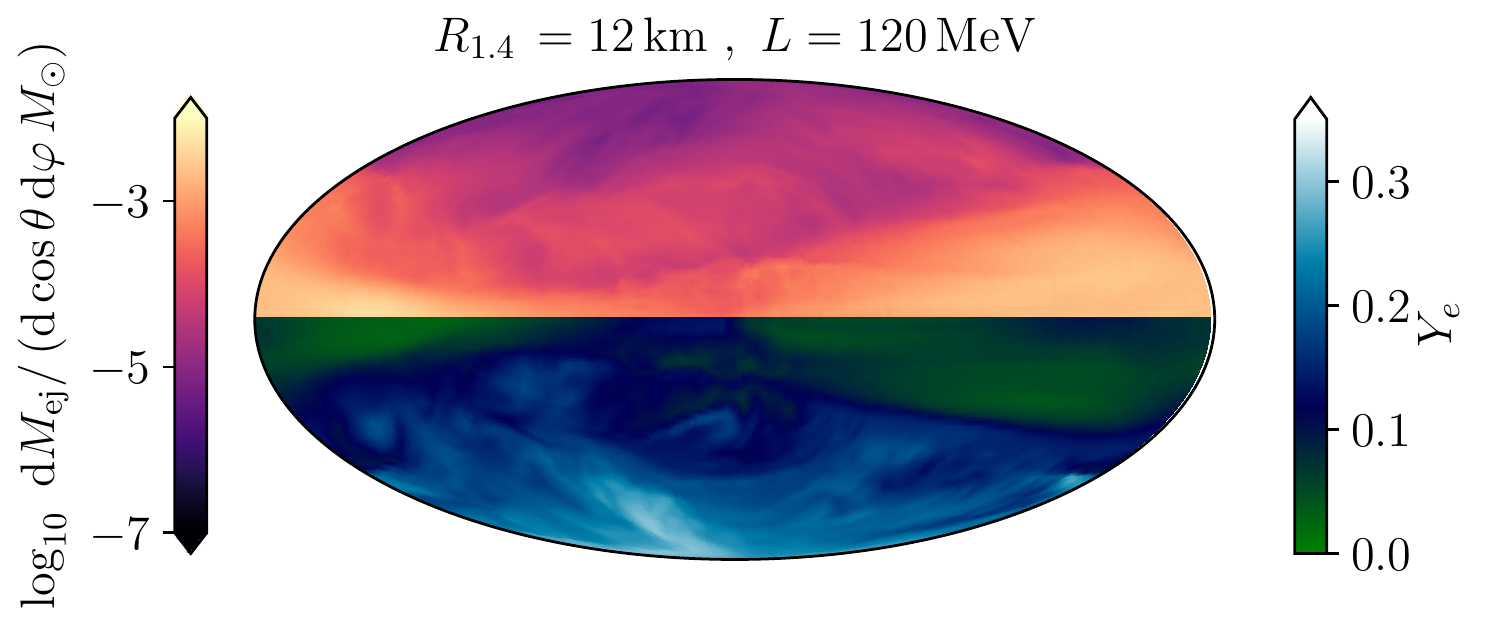}
  \includegraphics[height=3cm, trim=60 0 0 0, clip]{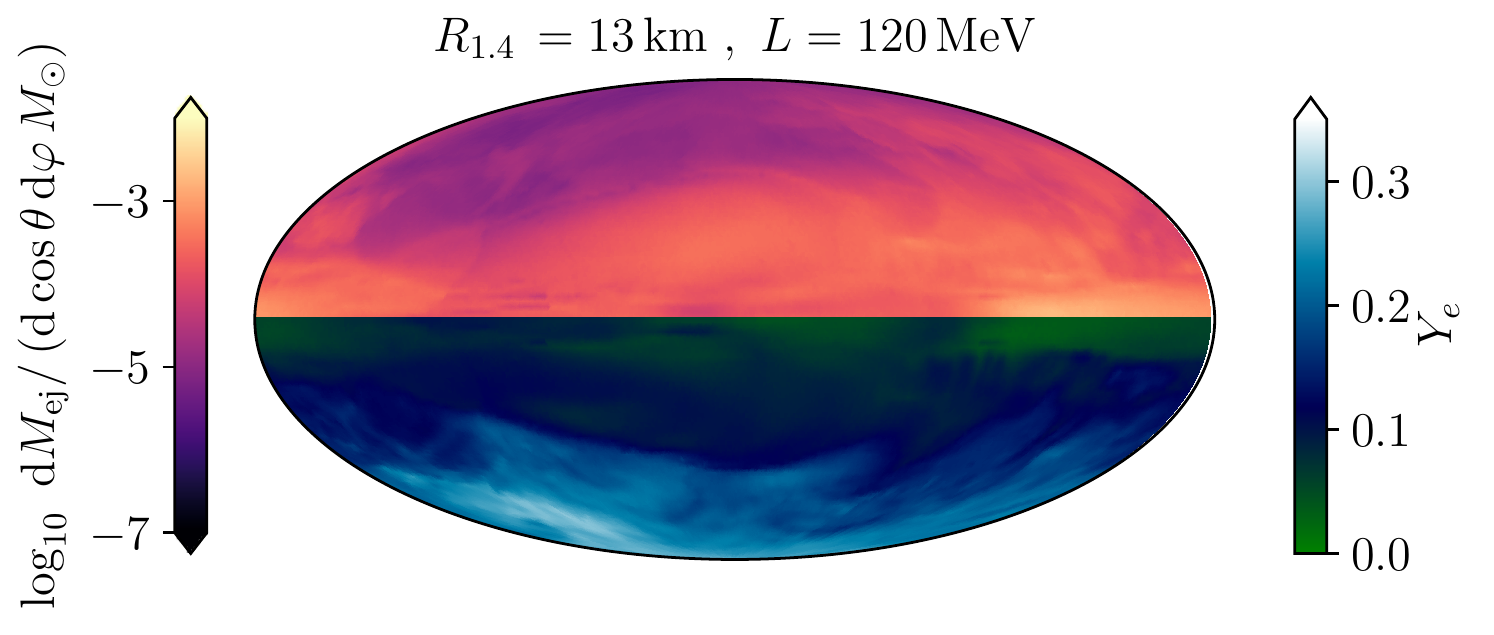}
  \caption{Time-integrated ejected mass $M_{\rm ej}$ and mass-weighted
    average electron fraction $Y_e$ projected onto a sphere at radius
    $r=295\,\rm km$ from the origin. The data is shown using Mollweide
  projection for the unequal mass models. }
  \label{fig:Mej_2D}
\end{center}
\end{figure*}

In this work, we extract the amount of mass ejecta on a spherical detector
placed at a radial coordinate $r=295\, \rm km$ from the merger site. We record the mass flux
\begin{align}
  \dot{M}_{\rm ej} = \oint_{r\,=\, 295\,\rm km} \sqrt{\gamma} \rho u^i \,
  {\rm d}S_i\,,
  \label{eqn:Mdot}
\end{align}
temperature $T$ and electron fraction $Y_e$ of the ejecta as the pass
through the surface $S$. Here $\rho$ is the rest-mass density, $u^i$ the spatial component of the fluid four-velocity, and $\gamma$ is the determinant of the 3-metric.
In order to determine whether a fluid element crossing
the detector is unbound, we use the $u_t <-1$ criterion \cite{2017CQGra..34u5005B}. We point
out that this will slightly underestimate the amount of unbound ejecta, as
it neglects the contribution of the internal energy.\\

We then time integrate the the mass flux to compute the amount of mass
ejection ${\rm d}M_{\rm ej}/{\rm d}\Omega$ per solid angle. Additionally,
we also compute the mass-weighted electron fraction as an indicator for
average nuclear composition.

The resulting spatial and
compositional distributions of the ejecta for the unequal mass mergers ($q=0.85$) are shown in Fig. \ref{fig:Mej_2D} using
Mollweide projection. 
Starting with the reference case of $R_{1.4}\,=\,12\,\rm km$, we observe the
following differences between the small $L\,=\,40\,\rm MeV$ case and the
$L\,>\,100\,\rm MeV$ cases. First, the $L\,=\,40\,\rm MeV$ case features
a rather spatially isotropic distribution of mass ejecta. Additionally, the electron
fraction reaches the highest average values in all three values of $L$, having
$Y_e>0.25$ for large parts of the mass ejection, for the $R_{1.4}\,=\,12\,\rm km$ EoSs.
With increasing slope parameter $L$, we find that the ejection becomes more
equatorial, with the largest amounts of ejecta in the $L\,=\,120\, \rm MeV$
case. Consistent with the increase of equatorial ejection, which is likely
tidally driven \cite{Sekiguchi:2015dma}, the electron fraction of the ejecta decreases to values below $Y_e< 0.1$ in those regions. Overall the electron fraction
reaches lower values also in polar regions for large $L$, compared to the $L\,=\,40\,\rm MeV$ cases. 
For unequal mass mergers, these trends are inversely correlated with $\Lambda_{1.4}$, whereas for equal masses the trend is less clear. 
From Fig. \ref{fig:MR}, we can see that there are strong variations in radius of the secondary star ($M_2=1.25\, M_\odot$ for the unequal mass binary).
As a result, we find that for those systems, tidal effects take over that correlate more strongly with the compactness. For those mass ratios, the larger $L$ models tend to produce more equatorial ejetca, despite having smaller tidal deformabilities.
Albeit somewhat counter intuitive, there have been previous examples in the literature in which tidal disruption was better captured in terms of the compactness, than with the tidal deformability \cite{Foucart:2018rjc}. These results suggest that perhaps the picture is more complicated than either a single compactness or $\Lambda$ parameter can generically capture.
Qualitatively, the same behaviour also applies to the $R_{1.4}\,\simeq\, 11\,\rm
km$ cases (Fig. \ref{fig:Mej_2D}, left column), where the $L\,=\,100\,\rm MeV$
simulation features enhanced neutron rich outflows in the equatorial plane
compared with the $L\,=\,40\,\rm MeV$ case, despite the fact that these EoSs have an identical $\Lambda_{1.4}$. Finally, as we saw for the $R\,=\,12\,\rm km$
models, the ejection in the $R_{1.4}\,=\,13\, \rm km$ cases is very similar for $L$=100~MeV compared to 120~MeV.\\
\begin{figure*}[!ht]
\centering
\includegraphics[width=\textwidth]{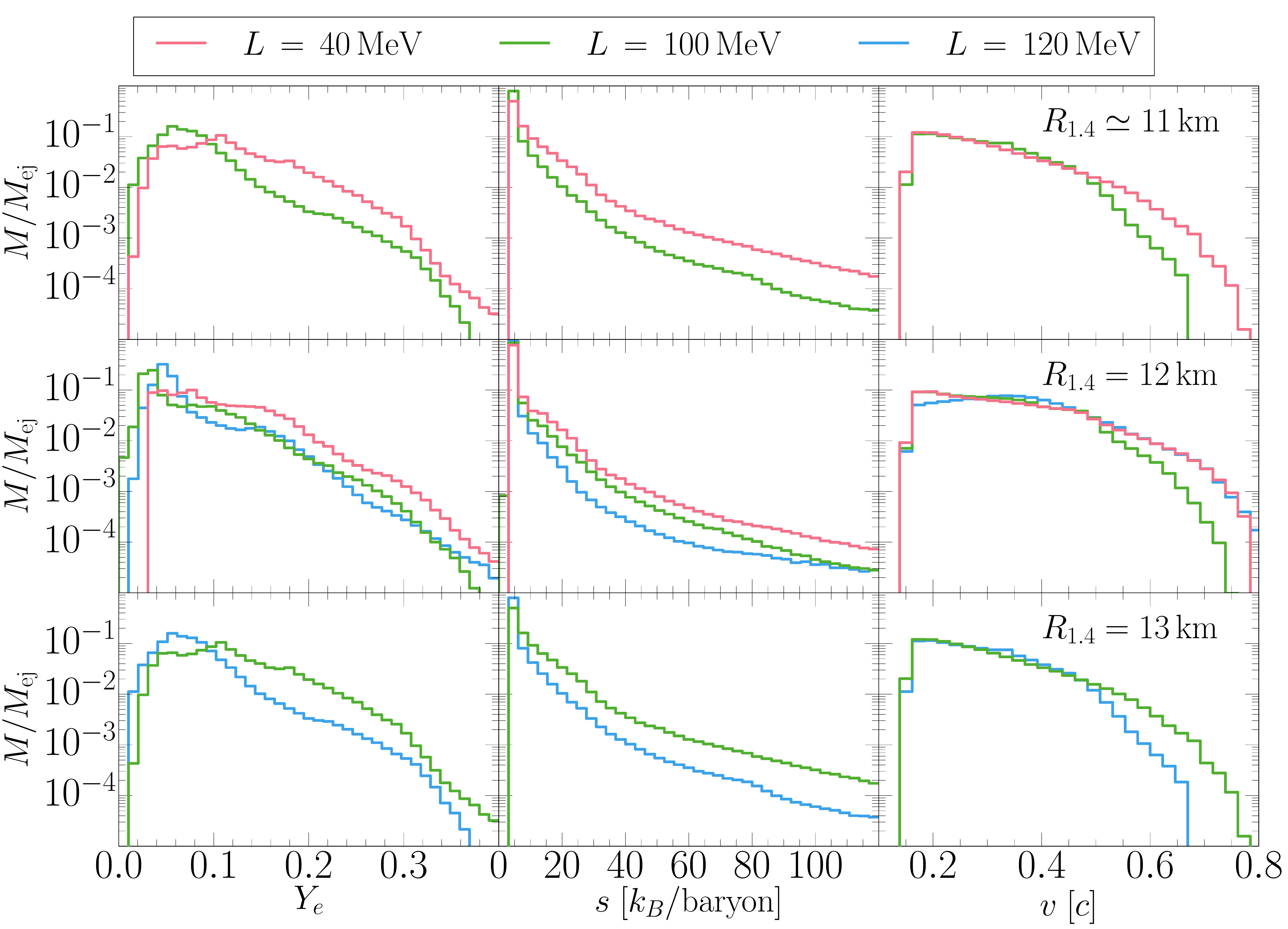}
\caption{\label{fig:ej_hist} Histograms of the electron
fraction, the average entropy per baryon, and the velocity of the
dynamical mass ejecta for simulations with mass ratios $q=0.85$.
The histograms refer to time integrated quantities and have been normalized to the respective amount of total mass
ejection $M_{\rm ej}$. The different colours refer to simulations with
different slopes $L$ of the nuclear symmetry energy, while $R_{1.4}$ denotes the radius
a $1.4\,M_\odot$ star for each EoS. }
\end{figure*}

In order to provide a more quantitative description of the mass ejection,
we next consider one-dimensional histograms of the entropy per baryon, $s$ , electron fraction, $Y_e$,
and velocity, $v$, \footnote{We estimate the velocity from the local Lorentz factor of the fluid element.} for the dynamical ejecta. These are shown in Fig.
\ref{fig:ej_hist}. Starting with the average electron fraction we can see
the overall distributions for the $R_{1.4}\,=\,12\,\rm km$ EoSs are surprisingly
similar. As discussed for Fig. \ref{fig:Mej_2D}, there are differences for
the lowest $Y_e$ bins, with large $L$ models containing slightly more mass at small $Y_e<0.05$, but the fall-off at large $Y_e$ is nearly identical for all values of $L$. This behaviour is very similar for the $R_{1.4}\,\simeq\,11\,\rm km$
cases.
Interestingly, the difference
between the $L\,=\,100\,\rm MeV$ and $L\,=\,120\,\rm MeV$ at intermediate
electron fractions $Y_e>0.15$ is more pronounced for large stars with
$R_{1.4}\,\simeq\,13\,\rm km$; however, these differences remain small. \\

Different from essentially all previous studies (see e.g. Refs. \cite{Baiotti:2016qnr,Dietrich:2020eud,Radice:2020ddv} for a review), our EoS are
specifically constructed to vary only in the high-density part, while also
using the same finite-temperature model (see Sec. \ref{sec:EOS}).
In all cases, the low density EoS is, however, the same. Given the
large insensitivity of the results to changes in high-density physics
between the models, this leads us to conjecture that the composition of the
ejecta is largely determined by low density EoS, which governs the outer regions of the stars from which they are ejected. 

In contrast, when considering the distribution of entropies $s$ per baryon for the mass
ejecta, we find a small trend with $L$. Specifically, we find that in all cases an ordering is present, where larger $L$
slope parameters can lead to a suppression of highly shocked material with
large, $s\,>\,40\, \rm k_B\,/\,baryon$. 

Finally, we comment on the prospect for high velocity ejecta, which is especially relevant in
the context of the recently observed X-ray rebrightening of GW170817 \cite{Hotokezaka:2018gmo}.
We find that, in all cases, high velocity tails with $v>0.6\, c$ are present,
which constitute about $1\%$ of the overall mass ejecta.
Different from the specific entropy $s$, no concrete ordering in terms of
$L$ can be inferred from our data. For large stars (bottom row of Fig.
\ref{fig:ej_hist}), higher values of $L$ lead to a suppression of fast
ejecta. On the other hand, for $R_{1.4}\,=\,12\,\rm km$ models, large and small
values of $L$ produce almost identical distributions, except at very low
velocities. It, therefore, seems that the dependence of fast ejecta on
nuclear parameters is more complicated, as already anticipated in earlier works 
\cite{Metzger:2014yda,Hajela:2021faz,Nedora:2021eoj}.

\begin{table*}[!th]
\centering
\begin{tabularx}{0.7\textwidth}{@{\extracolsep{\fill}}cccccccccccc}
\hline \hline
 $R_{1.4}$~[km] & $q$ & $L$~[MeV] & $M^{\rm dyn}_{\rm ej} \,\left[10^{-3}\, M_\odot\right]$   & $\left<v\right>\left[c\right]$ & $\left<Y_e\right>$ &$ T^{\rm max}_{\rm mer}  \left[\rm MeV\right]$ & $ T^{\rm max}_{\rm final} \left[\rm MeV\right]$ \\
\hline 
						&&    40\footnote{Note: the $R_{1.4}\simeq11$~km, $L=40$~MeV binary undergoes a delayed collapse $\sim$15~ms after merger.}  	&2.2&0.34&0.10& 136 & 63 \\  	
  $R_{1.4}$ $\simeq$  11& 0.85   &  100 &5.3 & 0.36& 0.07& 120 & 66&\\ \hline

	  			  &&    40	  &0.6&0.30&0.12& 89 & 40& \\ 	
			  &0.85 &	  100     &2.5&0.30&0.08& 95 & 60  \\
				 &&    120	 &6.2&0.31&0.07& 85 & 68 \\  
  \vspace{-0.1cm} $R_{1.4}$ = 12  & &&&& && \vspace{-0.1cm} 	\\		 
				 &&    40	  &1.2&0.28&0.09& 97 & 46& \\ 	
	 		 & 1.0 &  100     &0.7&0.27&0.13& 106& 61&  \\
				 &&    120	 &0.8&0.31&0.13&  129 & 52& \\ \hline
				
					 &&    100     &1.9&0.28&0.09& 94& 50 &  \\	
  $R_{1.4}$ = 13  &  0.85 	&      120   & 2.1& 0.28& 0.08& 85& 46& \\ \hline
\hline
\end{tabularx}
\caption{Summary of remnant temperatures and the mass-averaged dynamical ejecta properties. $M_{\rm ej}^{\rm dyn}$ is the total amount of dynamical ejecta, while $\left<v\right>$ and $\left<Y_e\right>$ are the mass-averaged velocity and electron fraction of the dynamical ejecta, respectively. $ T^{\rm max}_{\rm mer}$ is the maximum temperature achieved at densities above $\ns$ at the time of merger and $ T^{\rm max}_{\rm final}$ is the maximum temperature in the remnant ($n>\ns$) at the end of our simulations.}
  \label{table:ejecta}
\end{table*}

\subsection{Gravitational waves}
\label{sec:GW}

Finally, we consider the gravitational wave (GW) emission from the various EoSs in our sample. While several previous studies have used the inspiral of GW170817 to constrain $L$ \cite[e.g.,][]{Zhang:2018bwq,Raithel:2019ejc,Tsang:2019jva,Essick:2021kjb}, the dependence of the post-merger GW signal on $L$ has never before been systematically explored. Consistent with the rest of this paper, we thus focus our analysis in this section on the post-merger GW emission. Details on the analysis methods are summarized in Appendix \ref{app:GW}.

\begin{figure*}[!ht]
\centering
\includegraphics[width=\textwidth]{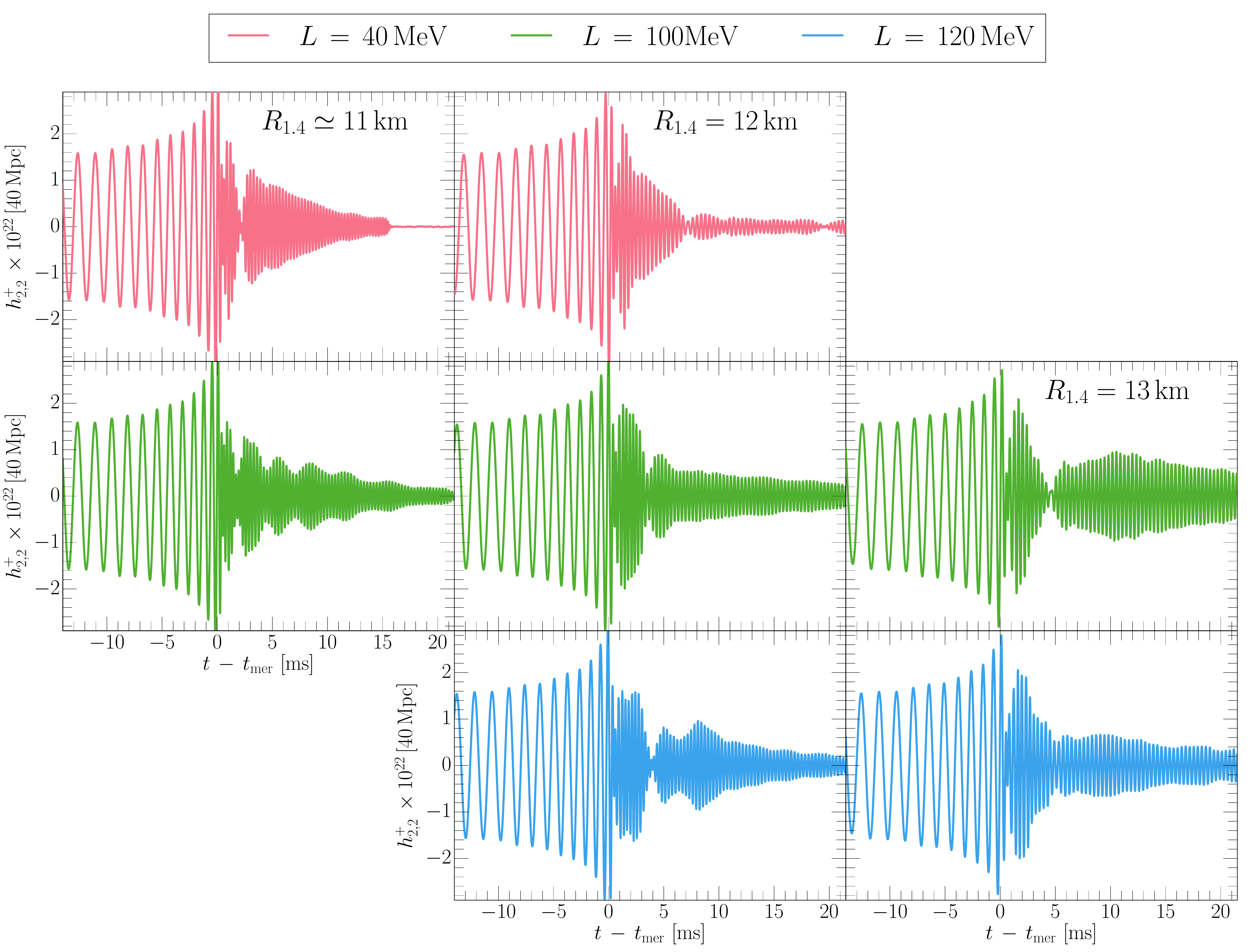}
\caption{\label{fig:strain_q085} Gravitational wave strain for the $q=0.85$ binaries, viewed face-on at 40~Mpc. The different radii are plotted in each column, while the rows show different values of $L$. All waveforms are aligned at the time of merger. We find significant differences in the post-merger GW signals, even for EoSs with the same $R_{1.4}$.}
\end{figure*}

\begin{figure*}
\centering
\includegraphics[width=0.9\textwidth]{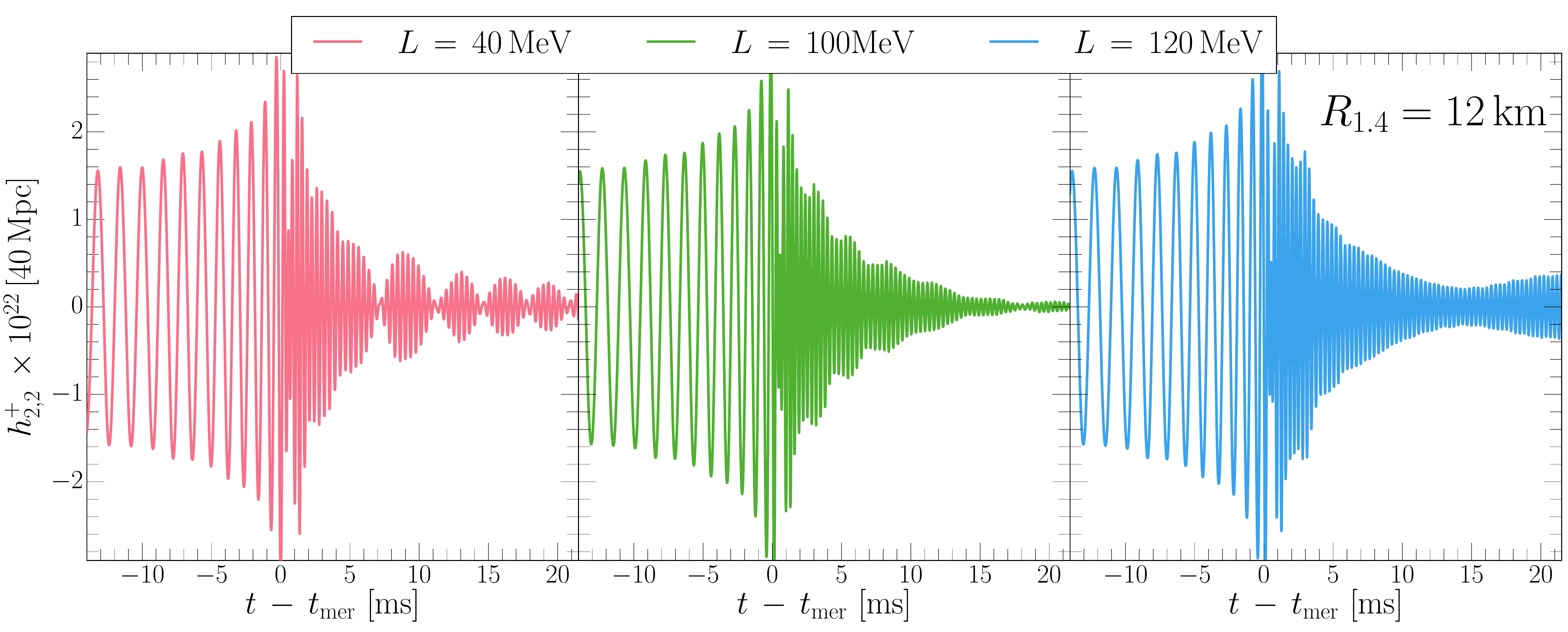}
\caption{\label{fig:strain_q1} Same as Fig.~\ref{fig:strain_q085}, but for the equal-mass binaries.}
\end{figure*}

\subsubsection{Gravitational wave signals}
\begin{figure*}[!ht]
\centering
    \includegraphics[width=0.9\textwidth]{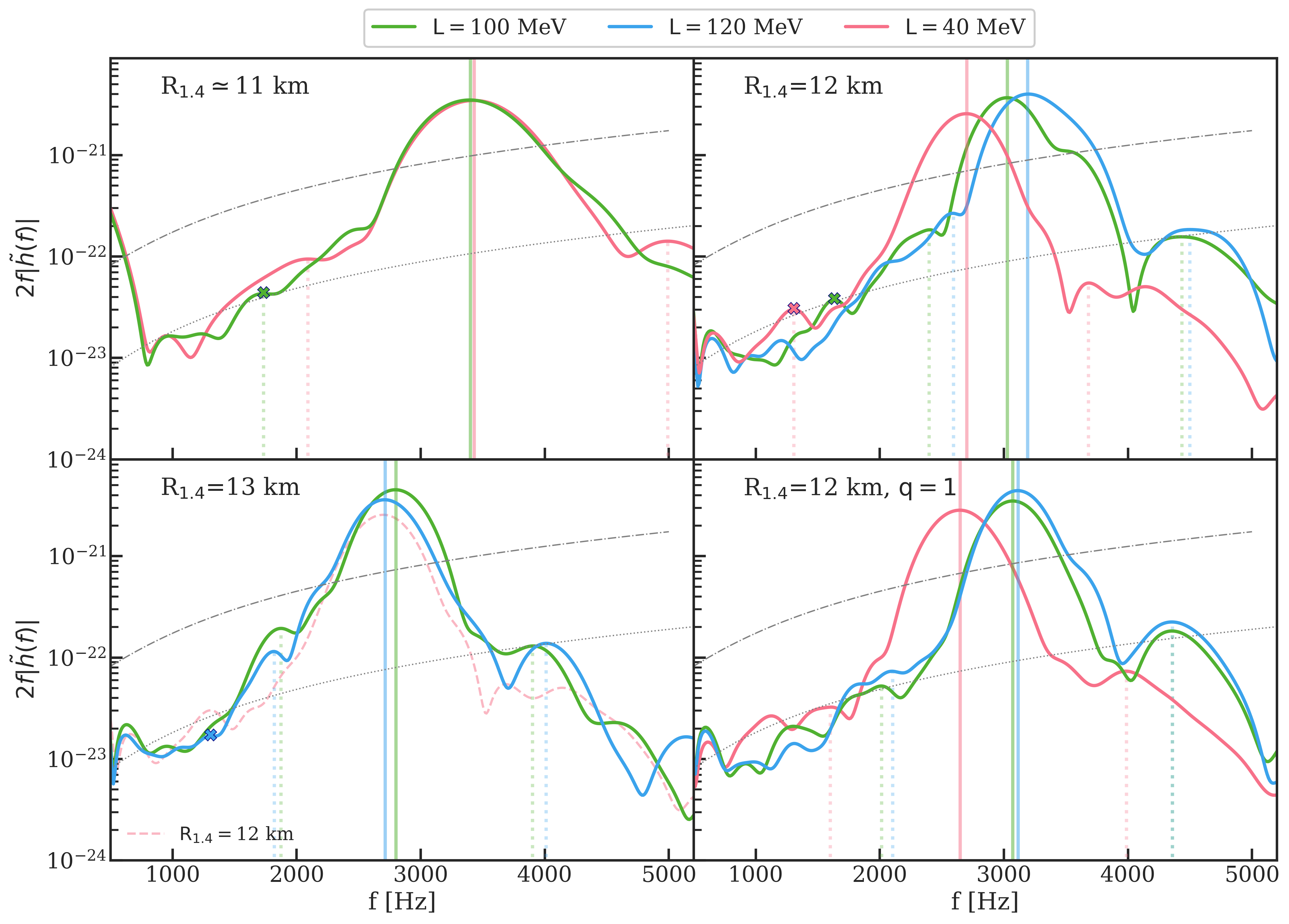}
\caption{\label{fig:hc} Characteristic strain, including all
$\ell$=2,~3 modes, for a face-on merger at 40~Mpc. The top two and bottom
left panels are for the $q=0.85$ binaries, while the bottom right
panel corresponds to the equal-mass binary.
The vertical solid line marks the
dominant $f_2$ spectral peak, while the dotted lines indicate the location of the
secondary peaks $f_1$ and $f_3$. The markers indicate the approximate location of spectral peaks associated
with the $m=1$ mode, which are expected to occur at $\simeq f_2/2$ where present. Finally, the
$R_{1.4}=12$~km, $L=40$~MeV EoS is also plotted in the lower left panel with the faded, dashed
pink line, to illustrate the similarity of this spectrum with the $R_{1.4}=13$~km spectra. The gray dash-dot and dotted lines indicate the design sensitivity curves for advanced LIGO \cite{LIGOScientific:2014pky} and Einstein Telescope \cite{Punturo:2010zz}, respectively.}
\end{figure*}
We start by showing the $\ell=m=2$ component of the plus-polarized GW strains, $h_{2,2}^+$, for the $q=0.85$ binaries in Fig.~\ref{fig:strain_q085}. These signals correspond to a face-on merger located at 40 Mpc. In all panels, we have aligned the signals at the time of merger, $t_{\rm mer}$, which we define as the time at which $\left|h_{+}^{2,2}\right|$ reaches a maximum. Although the EoSs are constructed with fixed radii $R_{1.4}$, their tidal deformabilities can differ significantly, see Table \ref{table:EoSs}. This leads to considerable phase difference of the waveforms by the time of merger. Only the EoSs with $R_{1.4}\simeq 11$~km have equal values of $\Lambda_{1.4}$, and these resulting binaries indeed have identical inspirals.

From Fig.~\ref{fig:strain_q085}, we observe significant differences in both the amplitude and beat frequencies of the decaying post-merger GWs for the various EoSs in our sample. We find differences between EoSs both with common $R_{1.4}=12$ or 13~km, and also between EoSs with identical tidal deformabilities (corresponding to the $R_{1.4}\simeq11$~km models, shown in the first column of Fig.~\ref{fig:strain_q085}). This suggests already that the values of $R_{1.4}$ or $\Lambda_{1.4}$ do not uniquely govern the post-merger GW emission. We also find differences in the post-merger GWs from the equal-mass binaries with $R_{1.4}$=12~km, which are shown in Fig.~\ref{fig:strain_q1}.

We explore the spectral content of these post-merger GWs in more detail by calculating their characteristic strain via eq.~(\ref{eq:hc}),
which we show in
 Fig.~\ref{fig:hc}, again for a face-on merger located at $40$~Mpc.  The spectra in Fig.~\ref{fig:hc}
show several well-defined peaks, which we highlight with vertical lines.
The dominant peak, which is marked with a solid vertical line and which
we call $f_2$, is located in each spectra at $\sim\,2800-3200$~Hz.   We also find
secondary peaks located to either side of $f_{2}$, which we call
$f_{1}$ and $f_3$ and which we mark with dotted vertical lines. Finally,
in a subset of the EoSs, we find a possible peak located at $\sim f_2/2$, which we call $f_{m=1}$ and we mark with a cross.
We summarize the location of all peaks in
Table~\ref{table:gw}.

We note that the $f_2$ spectral peak is typically associated with quadrupolar oscillations of the remnant, while the origin of
 the secondary peaks remains under debate \cite[][]{Stergioulas:2011gd,Takami:2014zpa,Baiotti:2016qnr, Paschalidis:2016vmz,Bauswein:2019ybt}. We do 
 not distinguish between the proposed origins in the present work, but rather treat the secondary peaks agnostically, 
 reporting simply the relative alignment of the peaks for each EoS. Additionally, we note that for some EoSs in our sample, these secondary peaks are only weakly resolved, and their exact alignment should be interpreted with a grain of salt. In the following, we, therefore, focus in particular on the dominant peak, $f_2$, and
 we return to a discussion of $f_{m=1}$ in Sec.~\ref{sec:onearm}.

In order to further quantify the differences of the gravitational waveforms in the post-merger phase, we compute detector-dependent overlap integrals, ${O}$, assuming the design sensitivity of Advanced LIGO \cite{LIGOScientific:2014pky} and a face-on source located at 40~Mpc (see Appendix \ref{app:GW} for details). The overlap integral is defined such that $O\lesssim0.992$ is required to marginally distinguish two waveforms with SNR of 8 \cite{Lindblom:2008cm,McWilliams:2010eq}.
For the $R_{1.4}\simeq 11$~km EoSs, we find that the post-merger GW signals are indistinguishable ($O=0.999$). The top left panel of Fig.~\ref{fig:hc} shows that location of the dominant spectral peaks are nearly identical as well (to within $\lesssim$30~Hz). These two EoSs have similar $R_{1.4}$ and identical $\Lambda_{1.4}=193$, yet differ substantially in $L$, ranging from 40 to 100~MeV. We plan to further explore this similarity of the GW emission for these two EoSs in a follow-up paper. We note, for now, that the similarity of these spectra suggests that $L$ does not have a clear imprint on the post-merger GW signal for this EoS.

The weak dependence of the post-merger GW signal on $L$ extends to larger values of $L$ as well. From the $R_{1.4}$=13~km EoSs, which have larger, albeit more similar, values of $L=100$ and 120~MeV, we find that the post-merger GWs are only marginally distinguishable ($O=0.97$) with Advanced LIGO at design sensitivity. Additionally, the maximum difference in $f_1$, $f_2$, and $f_3$ for these EoSs is $\lesssim 90$~Hz, suggesting again only a weak imprint of $L$ on the post-merger signal.

In contrast, for the $R_{1.4}$=12~km EoSs, we find a large difference between the GW signals for the $L=40$~MeV EoS compared to the EoSs with either $L=100$ or 120~MeV. These differences hold for both the equal and unequal-mass binaries. For the case of the unequal mass binary, the overlap integral between the $L=40$ and 100 (120)~MeV EoSs is 0.57 (0.34), while we additionally find differences of up to 490~Hz in the location of $f_2$ between these EoSs. The waveforms for these EoSs are thus clearly distinguishable for Advanced LIGO at design sensitivity. For the $R_{1.4}=12$~km EoSs with $L=100$ and 120~MeV, the overlap integral is still 0.88, indicating that even these signals can be distinguished with the sensitivity of Advanced LIGO. We find similar results for the equal mass binary, although for this case, the $L=100$ and 120~MeV spectra are only marginally distinguishable ($O=0.992$; see also Fig. \ref{fig:strain_q1}).

\begin{table}[!th]
\centering
\begin{tabularx}{0.5\textwidth}{@{\extracolsep{\fill}}cccccccccc}
\hline \hline
 $R_{1.4}$~[km] & $q$ & $L$~[MeV] & $f_1$ [kHz] & $f_2$ [kHz] & $f_3$ [kHz]  \\
\hline 
						&&    40\footnote{Note: the $R_{1.4}\simeq11$~km, $L=40$~MeV binary undergoes a delayed collapse $\sim$15~ms after merger.}  	& 2.09   &    3.43   &   4.99  \\  	
  $R_{1.4}$ $\simeq$  11& 0.85   &  100 & 1.73   &    3.40   &   7.29 & \\ \hline

	  			  &&    40	  &  1.31   &    2.70   &   3.68 \\ 	
			  &0.85 &	  100     & 2.40   &    3.03   &   4.43\\
				 &&    120	 & 2.59   &    3.19   &   4.50 \\  
  \vspace{-0.1cm} $R_{1.4}$ = 12  & & && \vspace{-0.1cm} 	\\		 
				 &&    40	  &  1.60   &    2.65   &   3.99 \\ 	
	 		 & 1.0 &  100     & 2.02   &    3.07   &   4.36 \\
				 &&    120	 & 2.10   &    3.12   &   4.36 \\ \hline
				
					 &&    100     & 1.87  &    2.80   &   3.90 \\	
  $R_{1.4}$ = 13  &  0.85 	&      120   & 1.82   &    2.71   &   4.01 \\ \hline
\hline
\end{tabularx}
\caption{Summary of post-merger GW frequencies. See the description in the text for further details.}
\label{table:gw}
\end{table}

\subsubsection{Correlation of $f_2$ with the high-density EoS}

\begin{figure*}[!ht]
\centering
\includegraphics[width=0.98\textwidth]{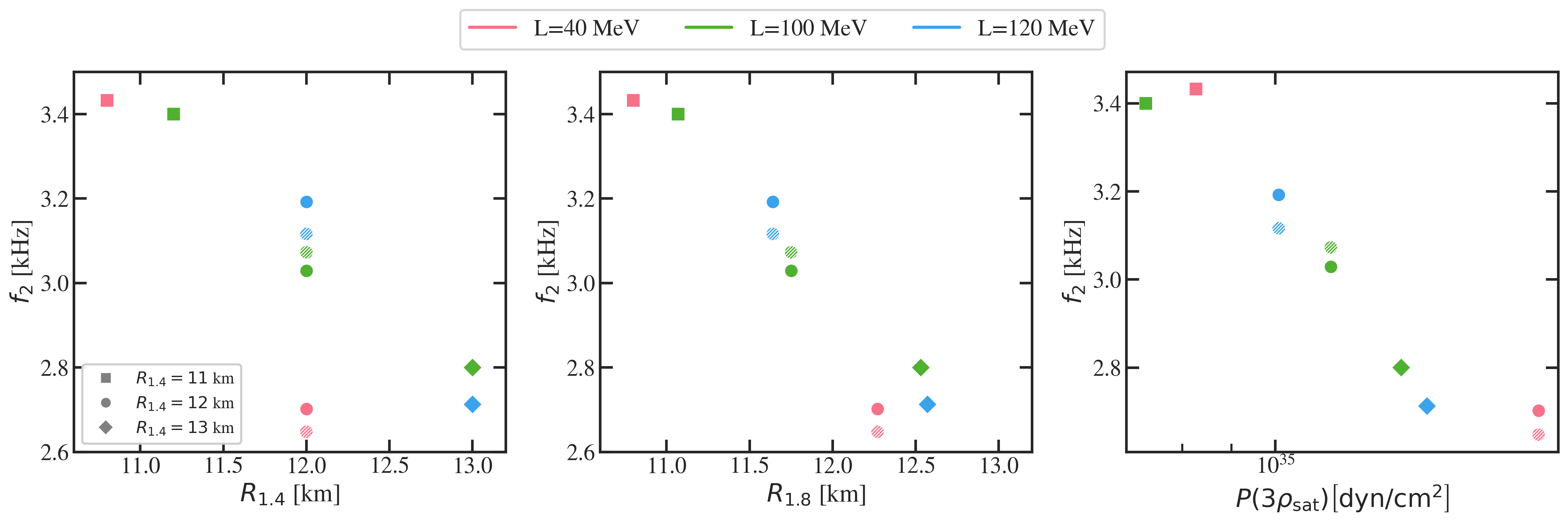}
\caption{\label{fig:fRP} Correlations between the peak frequency $f_2$ and the radii $R_{1.4}$ and $R_{1.8}$ of $1.4\,M_\odot$ and $1.8\, M_\odot$ neutron stars, respectively. We also show correlations with the pressure $P$ at density $n=3\,\ns$. The different colors correspond to the value of $L$, while the different symbol shapes indicate $R_{1.4}$ for our chosen EoSs. The hatched symbols correspond to the results from the equal-mass binaries, while the solid-fill symbols correspond to the $q=0.85$ binaries. We find a significant scatter in the correlation between $f_2$ and $R_{1.4}$, and that $f_2$ instead correlates more closely with the high-density part of the EoS, parameterized here either with $R_{1.8}$ or $P_{3 \ns}$. }
\end{figure*}
\begin{figure}[b]
\centering
\includegraphics[width=0.35\textwidth]{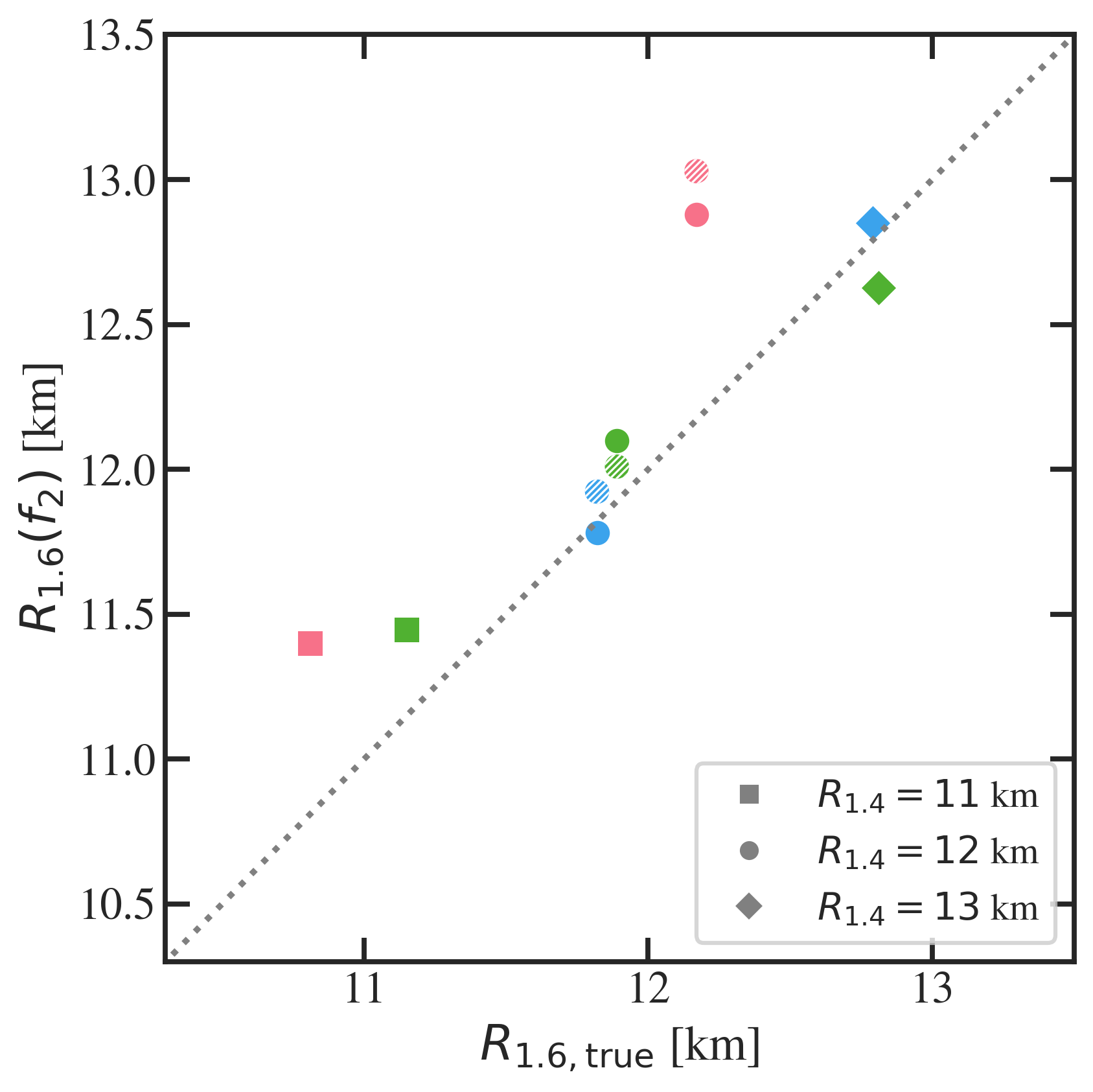}
\caption{\label{fig:RvsR} Inferred radius as a function of the true (inputted) radius, for each of the binaries in this work. The color scheme and symbols are as in Fig.~\ref{fig:fRP}. We calculate $R_{1.6}(f_2)$ using the universal relationship from eq.~(21) of \cite{Vretinaris:2019spn}. We find that using this standard fitting formula with these more extremal EoSs leads to inferred errors of up to 0.86~km in the radius for the EoSs included in this work.}
\end{figure}
\begin{figure*}[!ht]
\centering
\includegraphics[width=\textwidth]{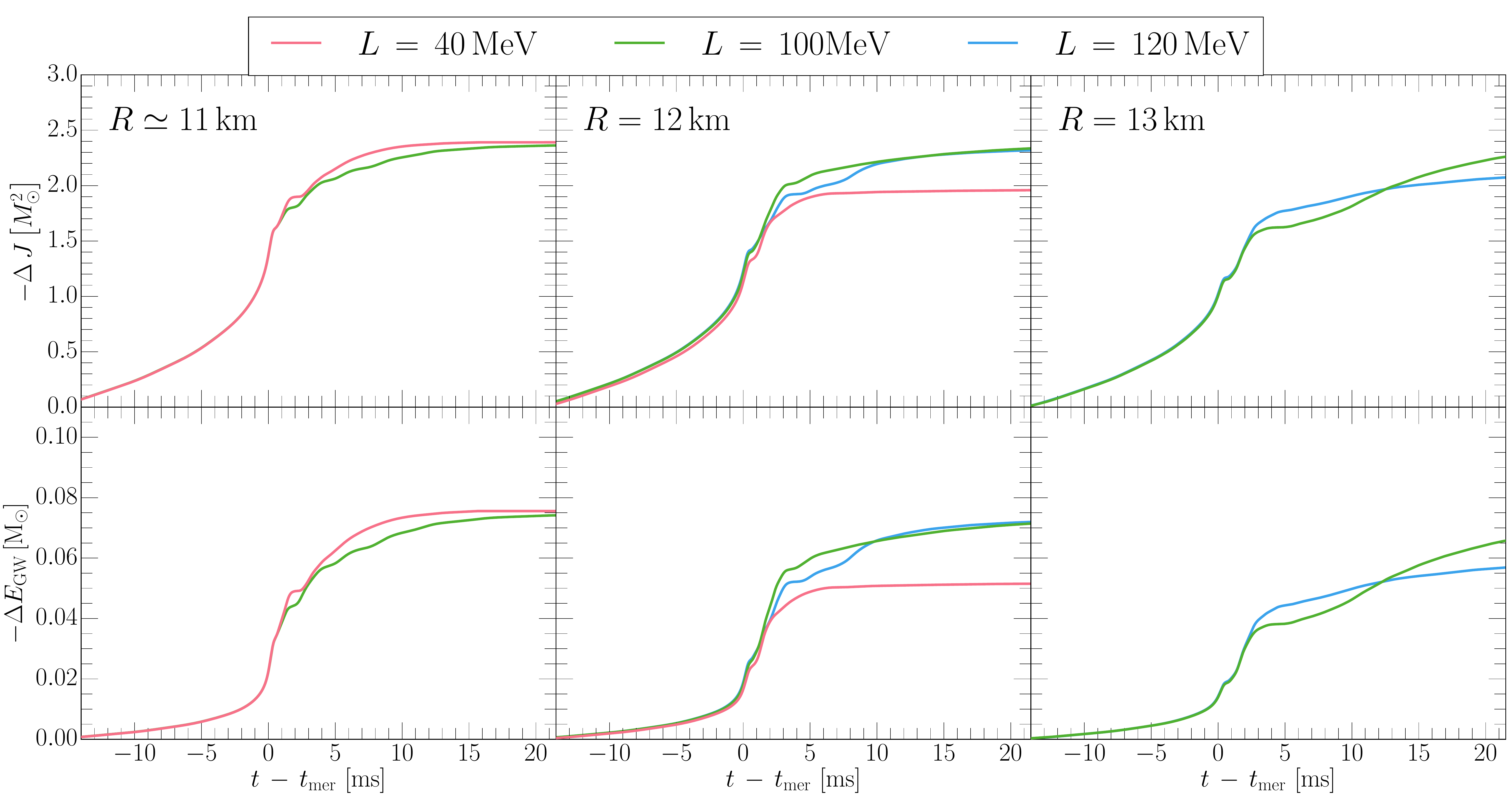}
\caption{\label{fig:EJ_q085} Gravitational wave energy ($\Delta E_{\rm GW}$) and angular momentum losses $\Delta J$ for the unequal mass mergers $q=0.85$. Different colors correspond to different slope parameters $L$.}
\end{figure*}
We therefore find that, although varying $L$ leads to significantly different GW emission for the EoSs with $R_{1.4}=12$~km, there is no clear trend between $L$ and the post-merger GWs that persists for all EoSs in our sample. Instead, we find that  $f_2$ correlates more strongly with the high-density EoS, as we show in Fig.~\ref{fig:fRP}. The three panels of Fig.~\ref{fig:fRP} shows $f_2$ as a function of $R_{1.4}$, $R_{1.8}$, and the pressure at $3\times$ the nuclear saturation density, $P(3\ns)$. We find differences of up to $\sim$500~Hz in $f_2$ for $R_{1.4}=12$~km. In contrast, the scatter in $f_2$ is substantially reduced by instead plotting against $R_{1.8}$ or $P(3\ns)$. In other words, we find a stronger correlation between $f_2$ and either $R_{1.8}$ or $P(3\ns)$, than with $R_{1.4}$. This is consistent with previous studies that have also found that $f_2$ correlates better with $R_{1.6}$ than with $R_{1.4}$ or even $R_{1.8}$ \cite{Bauswein:2012ya,Vretinaris:2019spn}. We find a similar strength of correlation with $R_{1.6}$ (not shown) as with $R_{1.8}$.

While $L$ is set by the pressure at $\ns$ (eq.~\ref{eq:PofL}), the pressure near $3-4~\ns$ primarily governs the slope of the mass-radius curve \cite{Ozel:2009da}. We thus find that $f_2$ may, in fact, be sensitive to the slope of the mass-radius curve. While the number of simulations performed here is insufficient to provide new fitting formulae for $f_2(P_{3\ns})$, these correlations suggest that $f_2$ may be able to probe the higher-density part of the EoS more cleanly than it probes $R_{1.4}$.

This dependence on the high-density pressure also explains the large scatter in $f_2$ for the $R_{1.4}$\,=\,12~km EoSs. In order for an EoS to have $L\,=\,40$~MeV, the pressure at $\ns$ must be relative soft. For that EoS to still reach $R_{1.4}=12$~km, it must undergo a rapid stiffening of the pressure, which in turn predicts larger radii for high-mass neutron stars (as shown in Fig.~\ref{fig:MR}). The $R_{1.4}$=12~km, $L=$40~MeV EoS actually has a similar high-mass radius to the $R_{1.4}$=13~km EoSs (as can be seen in Fig.~\ref{fig:MR}, and in Table~\ref{table:EoSs}). The similarity of $R_{1.8}$ for these different EoSs results in very similar spectra, as shown in the bottom left panel of Fig.~\ref{fig:hc}, where we overlay the $R_{1.4}=12$~km, $L\,=\,40\rm MeV$ spectrum (dashed, pink line) against the $R_{1.4}$=13~km spectra. We find that the $R_{1.4}=12$~km, $L\,=\,40\, \rm MeV$ EoS effectively masquerades as an $R_{1.4}=13$~km spectrum. In other words, based on the spectrum alone, one might infer that the GWs from the $R_{1.4}=12$~km, $L=$40~MeV EoS actually corresponds to a 13~km EoS, resulting in a 1~km error.

We can see this potential for error more clearly in Fig.~\ref{fig:RvsR}, in which we plot the true (inputted) radius against the radius inferred from $f_2$, which is calculated using the universal relations of Ref.~\cite{Vretinaris:2019spn}. We show the relationship for $R_{1.6}\left(f_2\right)$, which was found in that work to produce smaller residuals than $R_{1.4}\left(f_2\right)$ or $R_{1.8}\left(f_2\right)$. Indeed, we find slightly smaller residuals between the true and inferred radii when comparing with $R_{1.6}$, than with the fit formulae for $R_{1.4}$. Nonetheless, we still find errors of up to 0.86~km in the inferred radius, with the largest residual corresponding to the $R_{1.4}=12$~km, $L=40$~MeV EoS in the equal-mass configuration.

Understanding how to minimize the scatter in the relationship between $f_2$ and the $R$ is of critical importance if we are to achieve the long-standing goal of constraining the neutron star EoS with post-merger GWs \cite{Baiotti:2016qnr,Paschalidis:2016vmz,Bauswein:2019ybt,Bernuzzi:2020tgt,Radice:2020ddv}. For the Advanced LIGO/Virgo network operating at design sensitivity, the statistical measurement uncertainty on the radius inferred from $f_2$ may be as small as 100~m, for a merger at 20~Mpc ~\cite{Chatziioannou:2017ixj}. The error budget of such a measurement is thus likely to be dominated by the systematic uncertainty of the $f_2-R$ universal relationship \cite{Chatziioannou:2017ixj,Breschi:2019srl}. The number of detections required to constrain the radius to such an accuracy may also depend on the stiffness of the EoS \cite{Bose:2017jvk} and the distance of the source \cite{Haster:2020sdh}.
Of the models explored in this work, the largest scatter in this relationship comes from the $R_{1.4}=12$~km, $L=40$~MeV EoS. This EoS is a particularly interesting example, as it exhibits an extreme stiffening in the pressure and, accordingly, has a characteristic back-bend in the mass-radius relation (see Fig.~\ref{fig:MR}). Such EoSs are not commonly included in the simulations used to fit for the various $f_2-R$ universal relations reported in the literature \cite{Takami:2014zpa,Vretinaris:2019spn}. Rather, those samples tend to be dominated by EoSs with more vertical mass-radius relations. Our findings thus provide additional motivation to continue to systematically expand the library of EoSs used in neutron star merger simulations to explore a wider range of EoS phenomenology, in order to better quantify the uncertainties in the $f_2-R$ universal relations.

\subsubsection{Energy- and angular momentum loss}
We next focus on the amount of of energy, $\Delta E_{\rm GW}$, and angular momentum, $\Delta J$, carried away by gravitational wave emission. Understanding this loss of angular momentum, and in particular how quickly the neutron star remnant symmetrizes, has profound implications on the long term stability of the remnant. Small gravitational wave losses would aid a long lifetime of the (hyper-)massive neutron star, while very efficient emission could lead to an early collapse \cite{Shibata:2019ctb,Nathanail:2021tay}. It is interesting to ask whether changes in $L$ affect this emission, thus, leading to possible imprints of nuclear parameters onto the remnants life time.  

To this end, we show these losses for our unequal mass mergers ($q=0.85$) in Fig. \ref{fig:EJ_q085}. Focusing first on the $R_{1.4}\,=\,12\, \rm km$ cases, we find that the $L\,=\,40\,\rm MeV$ merger leads to the least amount of energy and angular momentum loss, with the emission essentially shutting off after $5\, \rm ms$ post-merger.
This indicates an extremely fast axisymmetrization  and a suppression in GW luminosity compared to previously studied EoSs \cite{Zappa:2017xba},  and also differs from the slow, but continued emission for all other EoSs in this work.
For example, this trend with $L$ does not hold up for other radii, with the $L\,=\,40\,\rm MeV$ system featuring the largest amount of emission at $R_{1.4}\,=\,11\, \rm km$. In the $R_{1.4}\,=\,13\, \rm km$ we do not find a clear trend with of $L$, with the $L=120$~MeV EoS leading to larger losses at early times and the $L=100~MeV$ EoS leading to larger losses at late times.

This strongly hints that the behavior of the post-merger remnant, which is governed by densities of several times $\ns$, is no longer strictly correlated with the behavior at $n_{\rm sat}$, and hence $L$. This is also consistent with the correlations we found between the $f_2$ frequencies and the high-density EoS in Fig. \ref{fig:fRP}.

\subsubsection{One-arm instability}
\label{sec:onearm}
Finally, we comment on the presence of a one-arm ($m=1$) spiral instability in the remnants. If saturated, this instability can lead to the development of a dense core that is offset from the remnant's center of mass. The resulting $m=1$ deformation in the density distribution drives the production of $(\ell,m)=2,1$ GW modes, which, in turn, generate a spectral peak located at $\sim f_2/2$ \cite{Paschalidis:2015mla}. First identified in the context of binary neutron star merger simulations in \cite{Paschalidis:2015mla,East:2015vix,East:2016zvv}, this instability has since been studied for a range of EoSs and binary configurations \cite{Lehner:2016wjg,Radice:2016gym}. It also also been suggested, that the continued presence of $m=1$ instability can inject energy into the disk and aid the production of spiral-wave winds on longer timescales \cite{Nedora:2019jhl}.

In Fig.~\ref{fig:hc}, we find a clear, albeit weak, peak at $f_{m=1}\approx f_2/2$ in the spectra for some of the $R_{1.4}=12$ and 13~km EoSs, for the unequal-mass binaries. We mark the location of these peaks with an ``x". For the $R_{1.4}=11$~km EoS, we do not find any peak within 10\% of $f_2/2$, although this may be a result of the lower resolution in these spectra. 

In order to further study the development of the $m=1$ mode, we show in Fig.~\ref{fig:Clm} the $(\ell,m)=2,1$ and $(\ell,m)=2,2$ modes of $\psi_4^{\ell,m}(t)$ (see Appendix~\ref{app:GW} for details). Figure~\ref{fig:Clm} is shown for an edge-on merger, which enhances the visibility of the $m=1$ mode  \cite[e.g.,][]{East:2015vix,East:2016zvv} and focuses again on the $R_{1.4}$=12~km EoSs. While the $m=1$ mode is subdominant to the $m=2$ mode in all cases, we find clear evidence of the one-arm instability developing at merger for each value of $L$, as evidenced by the rapid rise in $\psi_4^{2,1}$. Similar results are found for $R_{1.4}\, \simeq\, 11\, \rm km$ and 13~km EoSs. For the $q=1$ binaries, this mode quickly decays following merger. In contrast, for the unequal-mass binaries, the $m=1$ mode tends to saturate shortly after merger and remains persistent, even as the $m=2$ mode fades. The exception to this trend again comes from the $R_{1.4}=12$~km, $L=40$~MeV EoS.

\begin{figure}[!ht]
\centering
\includegraphics[width=0.45\textwidth]{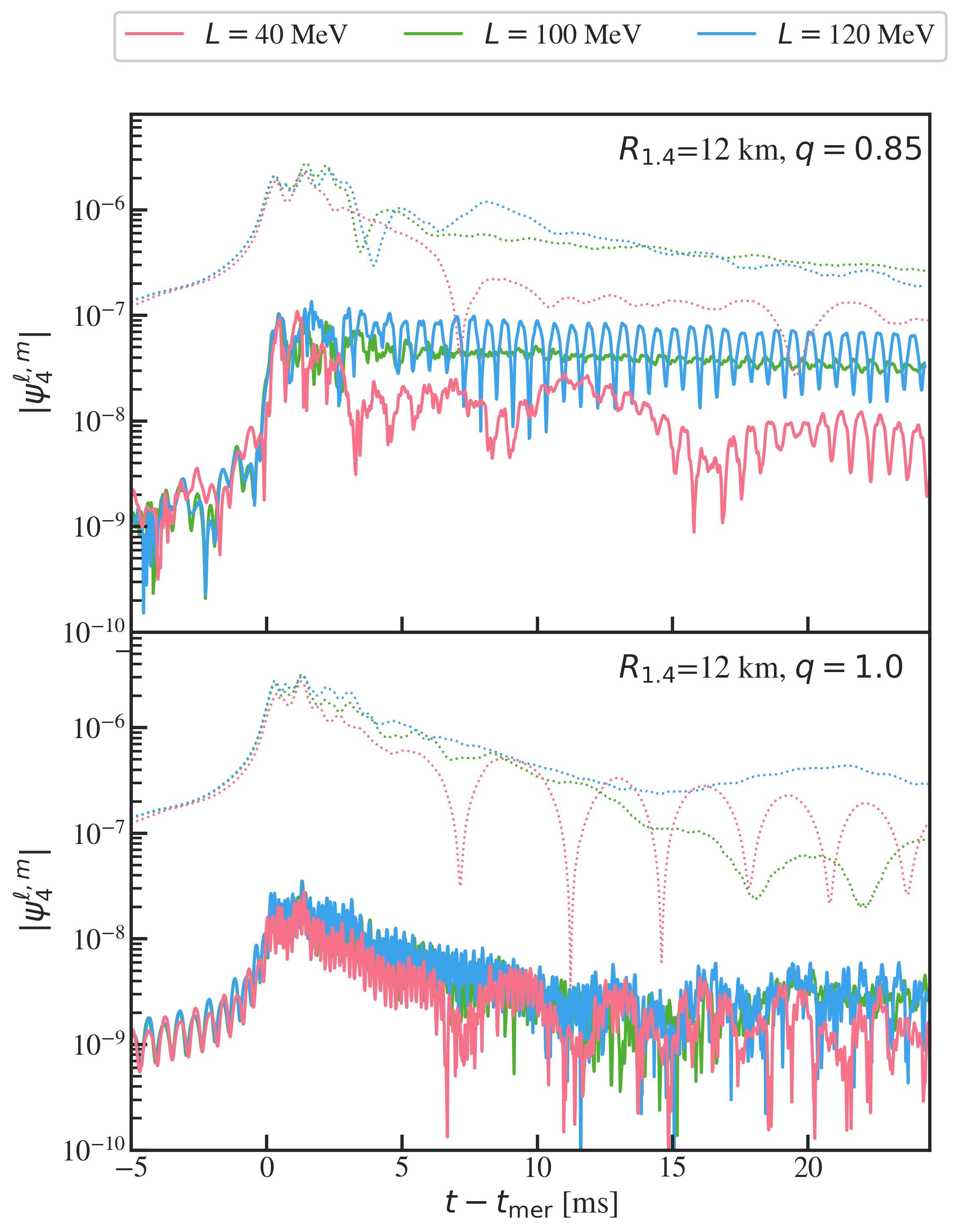}
\caption{\label{fig:Clm} Amplitude of the spherical harmonic components of the GW signal for an edge-on merger at 40~Mpc, for the $R_{1.4}=12$~km EoSs.
The solid and dotted lines correspond to the $(\ell,m)=(2,1)$ and (2,2) modes, respectively. For the $q=0.85$ binary (top panel), we find that the $m=1$ mode saturates within a few milliseconds of merger for the values of $L > 100$~MeV, but that the $m=1$ mode quickly decays for the $L=40$~MeV EoS, indicating that this binary is rapidly becoming axisymmetric. In contrast, for the equal-mass binaries (bottom panel), the $m=1$ mode is in general weaker, and decays quickly for all values of $L$.}
\end{figure}

For the $R_{1.4}=12$~km, $L=40$~MeV EoS, Fig.~\ref{fig:Clm} shows that the $m=1$ mode decays more quickly than any of the other, unequal-mass binaries. This is matched by a more rapid decay of $\psi_4^{2,2}$  in the same figure, as well as in the overall decay of the GW strain seen in Fig.~\ref{fig:strain_q085}. This damping of the GW emission suggests that the remnant quickly becomes axisymmetric. Indeed, the density contours in Fig.~\ref{fig:2D_temp} show that the $R_{1.4}=12$~km, $L=40$~MeV EoS is nearly axisymmetric, while the $L=100$ and 120~MeV EoSs still show a strong $m=1$ deformation at late times. We conjecture that this rapid axisymmetrization may be related to the extreme stiffening of this particular EoS. We leave further exploration of this finding to future work.

\section{Discussion and Conclusions}
In this work we have investigated the impact of systematically varying the slope $L$ of the nuclear symmetry energy on the post-merger dynamics, mass ejection, and gravitational wave emission of a binary neutron star coalescence. We have considered seven new EoSs, which were constructed to have $R_{1.4}\,\simeq\,11$~km (with identical $\Lambda_{1.4}$), or $R_{1.4}=\,12\,,13\, \rm km$, and to vary systematically in $L$ from 40 to 120~MeV.

Concerning the properties of the post-merger remnant, we have found that for our baseline models with $R_{1.4}\,=\,12\, \rm km$ varying the slope of $L$ significantly affects the temperatures probed in the remnant. In particular, we found that the EoS with $R_{1.4}=12$~km, $L=40$~MeV reaches lower temperatures $T\,\lesssim\, 40\,\rm MeV$ in the post-merger phase, whereas larger values of $L>100$~MeV for the same radius can reach  $T\,\gtrsim\, 60\,\rm MeV$. While the exact temperatures reached in the post-merger remnant are expected to also strongly depend on the total mass of the binary and the mass ratio, our results reported here are representative for a GW170817-like event and indicate a new sensitivity of the post-merger phase to the underlying EoS. 

This increase in temperature for large $L$ can be understood in terms of the compressibility of the EoS around $\ns$. High temperatures in a merger are only produced during the post-merger bounces of the neutron star cores. These bounces drive compressions of the outer layers of the stars and, as a result, should correlate strongly with the radii of the stars. To be more precise, we find that the temperatures probed during the merger itself are inversely proportional to $R_{1.4}$, whereas the late-time temperatures in the remnant are more correlated (still inverse-proportionally) to $R_{1.8}$. 
Same concern as in the results; this does not correlate well with R1.8.

We have also found that the post-merger remnant becomes more rapidly axisymmetric for the $R_{1.4}=12$~km, $L=40$~MeV EoS, compared to EoSs with identical $R_{1.4}$ and larger $L$. In particular, we found that the $m=1$ deformation that is naturally induced in an asymmetric merger, e.g. with mass ratio $q=0.85$ considered here, is rapidly dampened for this small $L$ EoS. We do not find evidence of a similar damping in any other EoS we studied, including the $R_{1.4}\simeq11$~km, $L=40$~MeV EoS; leading us to conjecture that the effect may in fact stem from the high-density EoS, rather than the specific value of $L$. Further studies with additional EoSs will be needed to clarify this behaviour. 

Since post-merger gravitational wave signals are very promising probes of the dense matter EoS \cite{Baiotti:2016qnr,Paschalidis:2016vmz,Bauswein:2019ybt,Bernuzzi:2020tgt,Radice:2020ddv}, we also performed a detailed analysis of the GW emission extracted from our simulations. Interestingly, we found that the $R_{1.4}\,=\,12\rm km$ EoSs exhibit significantly different post-merger GW emission for the three values of $L$, which may be distinguishable with Advanced LIGO at design sensitivity. In particular, we found that for the EoSs with $R_{1.4}=12$~km, the $L=40$~MeV and $L=100$~MeV EoSs differ in $f_2$ by more than 300~Hz for both equal- and unequal-mass binaries, and differ by nearly 500~Hz for the $L=40$~MeV EoS compared to the $L=120$~MeV EoS.

In contrast, for the $R_{1.4}\simeq11$~km EoSs, we found no significant differences in the post-merger GWs between the $L=40$ and 100~MeV cases, and we found only minor differences between the $R_{1.4}=13$~km EoSs, strongly hinting that $L$ is not uniquely imprinted in the gravitational wave signal.
More precisely, whereas $L$ governs the behavior of the EoS around $n_{\rm sat}$,  our findings suggest that the large differences in these EoSs at densities above $2\,n_{\rm sat}$ might be playing a stronger role in the dynamics of the post-merger system. 

To further illustrate this point, we compared the results from our post-merger GW spectra to known quasi-universal relations that relate the post-merger peak frequencies to the neutron star radius \cite{Vretinaris:2019spn} and found that using the standard relations to map from $f_2$ to the $R_{1.6}$ can lead to errors in the inferred radii of up to $\sim1$~km, for the EoSs considered in this work. As shown in Fig. \ref{fig:fRP}, we found that while the correlation between $f_2$ and $R_{1.4}$ is weak, a better result may be obtained by comparing $f_2$ with either the radius at higher masses, $R_{1.8}$, or by correlating $f_2$ with the pressure at $3\ns$. This underscores the importance of the high-density EoS, rather than $L$, in governing the post-merger GW spectrum. 

Finally, we also considered the dynamical ejection of matter during the merger. Since this material is ejected from the outer parts of the star, we expect to here find the strongest correlation with $L$, as $L$ affects the EoS around $\ns$. Indeed, or the $q=0.85$ binaries, we found that smaller values of $L$ lead to a systematic (and monotonic) reduction in mass ejection (Tab. \ref{table:ejecta}), with a corresponding, systematic increase in shock heating (Fig. \ref{fig:ej_hist}). However, this trend is less clear in the subset of equal-mass binaries. In contrast,  the compositional properties of the ejecta are similar in all cases and we only find minor differences in the fast tails of the ejecta velocities.
Although only indicative at this point, due to the limited set of EoS and mass ratios considered here, we plan to follow up on these potential trends with $L$ in future works.
This preliminary correlation between $L$ and the mass ejected may have implications on the feasibility of such systems to produce an X-ray and radio rebrightening years after the merger, as has recently been suggested for GW170817 \cite{Hajela:2021faz,Balasubramanian:2021kny}, and may provide a new pathway for constraining the slope of the nuclear symmetry energy from observations of future electromagnetic counterparts. Determining the feasibility of such constraints, and the precise dependence on the mass ratio, will be the focus of future work.

Thus, although the post-merger dynamics and GWs do not show a clear signatures of the slope of the nuclear symmetry energy, we find a potential new correlation between $L$ and several properties of the dynamical ejecta.
Our work also showcases the need for targeted EoS modelling, to systematically vary nuclear matter parameters while keeping the finite-temperature part of the EoS fixed, as was utilized in this study. Such targeted modelling will be crucial for better understanding systematic variations in the $f_2-R$ universal relations, as well as the differences in symmetrization timescales for post-merger remnants with different EoSs. Systematic construction of new EoSs will also allow us to follow-up on the tantalizing trends uncovered in this work, between $L$ and the properties of the dynamical ejecta, which may one day allow for new, astrophysical on the nuclear symmetry energy.

\begin{acknowledgements}
We thank L. Jens Papenfort for providing low density EoS matching routines.
ERM and CR gratefully acknowledge support from postdoctoral fellowships 
at the Princeton Center for Theoretical Science, the Princeton
Gravity Initiative, and the Institute for Advanced Study. 
CR additionally acknowledges support as a John
N. Bahcall Fellow at the Institute for Advanced Study.
This work used the Extreme Science and Engineering Discovery Environment (XSEDE), which is supported by National Science Foundation grant number ACI-1548562. 
The authors acknowledge the Texas Advanced Computing Center (TACC) at The University of Texas at Austin for providing HPC resources that have contributed to the research results reported within this paper, under LRAC grants AT21006 and AT20008.
Additionally, the authors are pleased to acknowledge that the work reported on in this paper was partially performed using the Princeton Research Computing resources at Princeton University which is consortium of groups led by the Princeton Institute for Computational Science and Engineering (PICSciE) and Office of Information Technology's Research Computing.
Part of this work was performed at the Aspen Center for Physics, which is supported by National Science Foundation grant PHY-1607611. The participation of ERM at the Aspen Center for Physics was supported by the Simons Foundation.
\end{acknowledgements}

\appendix

\section{Low-density symmetry energy}
\label{sec:Esym_lown}
At densities below 0.5$\ns$, nuclei start to form and the nuclear symmetry energy expansion formalism breaks down. Accordingly, we start the transition from our analytic, high-density EoSs to a tabulated low-density EoS at $n=0.5\ns$, as described in Sec.~\ref{sec:EOS}. However, across the transition window, between $6.3\times10^{-5}$ and 0.08~fm$^{-3}$, the total energy is given by a combination of the tabulated EoS and the analytic model. In this regime, we therefore need a reasonable extrapolation of the symmetry energy model to combine with the low-density EoS table. In this appendix, we describe the treatment of the symmetry energy in this low-density regime.

We empirically choose a power-law decay model for the symmetry energy extrapolation, to ensure that $E_{\rm sym}$ (1) remains positive and real, (2) provides a diminishing contribution to the overall energy, and (3) predicts $Y_{e,\beta} \in (0,0.5]$, with $Y_{e,\beta}$ approaching that of SFHo at low densities.  

For $n < 0.5\ns$, we thus adopt the following model for the symmetry energy,
 \begin{equation}
 \label{eq:Esymlow}
E_{\rm sym, low}(n) = [1-\chi(n)] E_{\rm fl} + \chi(n) E_{\rm PL}(n)
\end{equation}
where $E_{\rm PL}(n)$ is a power-law function and $E_{\rm fl}$ is an energy floor that we set to 11.22~MeV. This floor corresponds to $E_{\rm sym}(0.5\ns)/2$, calculated with the best-fit parameters for the SFHo EOS ($S=31.47$~MeV, $L=47.10$~MeV, $\gamma=0.41$; \cite{Raithel:2019gws}). In this expression, $\chi(n)$ is a smoothing function that we define as
\begin{equation}
\chi(n) = \frac{1 + \tanh\left[ X \left( n - n_0 \right)\right]} {2},
\end{equation}
where we choose $X=40$ and $n_0=0.025$~fm$^{-3}$, such that $\chi(n)\approx1$ at $0.5\ns$ (to within 1\% accuracy), and $\chi(n)$ decreases at lower densities. These parameters and the value of $E_{\rm fl}$ were empirically chosen to ensure $Y_{p,\beta}(n)$ approximately matches that of SFHo across this density regime, for the EoSs explored in this work.

In order to ensure continuity in the symmetry energy and the corresponding pressure, we define the power-law energy extrapolation according to
\begin{multline}
E_{\rm PL}(n) =  E_{\rm sym}(n_t) +  
	 	\frac{P_{\rm sym}(n_t)}{n_t (\gamma_{\rm PL}-1) } \left[ \left(\frac{n}{n_t}\right)^{\gamma_{\rm PL}-1} -1\right],
\end{multline}
where $n_t=0.5\ns$, the power-law index is given by
\begin{equation}
\gamma_{\rm PL} = \frac{\partial P_{\rm sym}(n) }{\partial n} \biggr \rvert_{n_t} \left[ \frac{n_t}{P_{\rm sym}(n_t) } \right],
\end{equation} 
and $P_{\rm sym}(n)=n^2 \partial E_{\rm sym}(n)/\partial n$.

In calculating the corresponding model for the low-density symmetry pressure, we neglect the density-derivatives of $\chi(n)$, which introduce unphysical density-dependences. Instead, we calculate the pressure in the two asymptotic limits, and use $\chi(n)$ to smoothly connect these regimes, i.e.,
\begin{align}
P_{\rm sym, low}(n) &=  n^2 \left( \frac{\partial E_{\rm fl}}{\partial n} \right) [1-\chi(n)] + n^2 \left( \frac{\partial E_{\rm PL}(n)}{\partial n} \right) \chi(n)  \\ \notag 
&= P_{\rm PL}(n) \chi(n)
\end{align}
where the first term disappears because $E_{\rm fl}$ is a constant and the remaining term is given simply by
\begin{equation}
P_{\rm PL}(n) =  P_{\rm sym}(n_t) \left(\frac{n}{n_t}\right)^{\gamma_{\rm  PL}}.
\end{equation}

\section{Chemical potentials}
\label{sec:mu}
In this appendix, we describe the calculation of the chemical potentials, which are used to determine the neutrino transport opacities within our numerical evolutions (following Appendix A of \cite{1996A&A...311..532R}). 

Because chemical potentials cannot be straightforwardly calculated within the original $M^*$-framework of Ref.~\cite{Raithel:2019gws}, we here introduce an approximate calculation for the chemical potentials. We take advantage of the fact that
the neutrino opacities depend primarily on the difference between the nucleon chemical potentials, 
\begin{equation}
\hat{\mu}(n,Y_p,T) \equiv \mu_n(n,Y_p,T)  - \mu_p(n,Y_p,T) ,
\end{equation}
where $\mu_n$ and $\mu_p$ are the neutron and proton chemical potentials, respectively.
The individual nucleon potentials, $\mu_n$ and $\mu_p$, do not enter the calculation of the absorption opacities, and they enter the scattering opacity only via a term that accounts for Pauli blocking among the degenerate nucleons \cite{1996A&A...311..532R}. Moreover, because Pauli blocking is relevant only at high densities, where the bulk of the matter is already expected to be optically thick to neutrino transport \cite[e.g.,][]{Endrizzi:2019trv}, we do not expect the exact treatment of $\mu_n$ and $\mu_p$ to significantly affect the final optical depth of the remnant, as long as $\hat{\mu}$ is modeled accurately. 

We, therefore, construct a self-consistent model for $\hat{\mu}$ at high-densities, based on the nuclear symmetry energy model used throughout this paper. We then define the individual chemical potentials relative to the values from the SFHo EoS table, such that $\mu_n(n, Y_p, T)=\mu_{n, \rm{SFHO}}(n, Y_p, T)$ and $\mu_p(n,Y_p,T) \equiv \mu_{n, \rm{SFHO}}(n, Y_p, T) - \hat{\mu}(n, Y_p, T)$. Because we are already matching to the full SFHo table at low densities, using this EoS provides maximum consistency in our calculations.
 We again stress that this is mostly done for convenience with respect to the existing numerical infrastructure. Except for Pauli-blocking, the values for $\mu_n$ and $\mu_p$ never enter separately in our simulations. Furthermore, $\beta-$equilibrium is solely determined by the difference, $\hat{\mu}$.

We calculate $\hat{\mu}$ from the symmetry energy as follows.
The chemical potential of species $i$ is defined as 
\begin{equation}
\label{eq:mu}
\mu_i \equiv \frac{ \partial E_i }{ \partial Y_i} \biggr \rvert_{S,n}
\end{equation}
where $E_i$ is the energy per baryon of that species, $Y_i$ is the corresponding number fraction, and $S$ is the entropy.
For uniform \npe matter, this implies
\begin{equation}
\label{eq:dEdYp}
\frac{\partial E_{\rm tot}(n,Y_p)}{\partial Y_p} = \mu_p + \mu_e - \mu_n,
\end{equation}
where $E_{\rm tot}$ represents the total energy, including contributions from neutrons, protons, and electrons. We have here assumed charge neutrality and conservation of baryon number, which require that $Y_e=Y_p$ and $Y_p = 1-Y_n$, respectively.
As in eq.~(\ref{eq:Enucl}), the total energy for zero-temperature \npe matter can also be written in terms of the symmetric matter energy and a symmetry energy correction, i.e.,
\begin{equation}
\label{eq:Etot_appendix}
E_{\rm tot}(n,Y_p,T=0) = E_0(n) + E_{\rm sym}(n) (1-2Y_p)^2 + E_e(n,Y_e),
\end{equation}
where we have additionally included the energy contribution from electrons, $E_e(n,Y_e)$. Differentiating with respect to $Y_p$, we find
\begin{equation}
\label{eq:dEtot_v2}
\frac{\partial E_{\rm tot}(n,Y_p,T=0)}{\partial Y_p} = -4(1-2Y_p) E_{\rm sym}(n) + \mu_e.
\end{equation}
Combining eqs.~(\ref{eq:dEdYp}) and (\ref{eq:dEtot_v2}), the zero-temperature difference in chemical potentials for neutrons and protons is then 
\begin{equation}
\label{eq:muhat}
\hat{\mu}(T=0) = 4 (1-2 Y_p) E_{\rm sym}(n).
\end{equation}
Equation~(\ref{eq:muhat}) thus ensures that $\hat{\mu}$ is consistent with the complete EoS model, for a given set of $E_{\rm sym}$ parameters.
We note that this approach assumes that the thermal part of the chemical potential is the same for neutrons and protons. While this is an approximation, it is consistent with the overall decomposition of thermal effects from composition-dependent effects in the EoS framework of Ref.~\cite{Raithel:2019gws}, where it was found that adding in the composition correction to the thermal model had a negligible effect on the total energy of the EoS (see \cite{Raithel:2019gws} for further discussion). 

Finally, we use the tabulated values from SFHo for the electron chemical potentials, which are simply given by the normal chemical potential for the Fermi-Dirac distribution function \cite{Hempel:2009mc}.\footnote{See also the EoS manual from the webpage of M. Hempel, https://astro.physik.unibas.ch/en/people/matthias-hempel/equations-of-state/.}

We reiterate that, on the short timescales ($\sim25$~ms) considered in this paper, the high-density matter is expected to remain optically thick to neutrinos \cite{Endrizzi:2019trv}. Thus, we do not expect this high-density approximation for the chemical potentials to affect the outcomes of our evolutions or any of the conclusions in this paper. However, for longer-term evolutions, for example to simulate cooling of the neutron star remnant, this approximation may not be sufficient and should be further tested before use.

\section{Details on the gravitational wave analysis}
\label{app:GW}
Finally, in this appendix, we detail our methods for analyzing the GW emission. We extract the GW signal from our simulations using the Newman-Penrose scalar $\psi_4$, which is related to the GW strain according to $\psi_4 = \ddot{h}_+ - i \ddot{h}_{\times}$, where $h_+$ and $h_{\times}$ are the plus- and cross-polarizations of the GW strain and the dots indicate derivatives with respect to time. We decompose $\psi_4$ into $s=-2$ spin-weighted spherical harmonics on spheres of large radius ($r=300~\Ms$), according to
\begin{equation}
\label{eq:psi4}
\psi_4(t'',r,\theta, \phi) = \sum_{\ell=2}^{\infty} \sum_{m=-\ell}^{\infty} \psi_4^{\ell,m}(t,r) _{-2}Y_{\ell,m}(\theta, \phi)
\end{equation}
where $t$ is the time and the angles $\theta$ and $\phi$ are defined with respect to the angular momentum axis. The total strain is then given by
\begin{align}
h(t) & \equiv h_+ - i h_{\times} \\ \notag
   & = \int_{-\infty}^{t} dt' \int_{-\infty}^{t'} dt'' \psi_4(t'',r,\theta, \phi),
\end{align}
where to calculate the double time integral, we use the fixed frequency integration (FFI) method of \cite{Reisswig:2010di}. 

In order to study the spectral features of the post-merger signals, we additionally calculate the characteristic strain, which is conventionally defined as
\begin{equation}
\label{eq:hc}
h_c(f) = 2 f \tilde{h}(f), 
\end{equation}
where $f$ is the frequency and $\tilde{h}(f)$ is the Fourier transform of $h(t)$
 \cite{Moore:2014lga} To calculate $\tilde{h}(f)$, we first window $\psi_4(t)$
 between $t_{\rm mer}+1.5$~ms and the maximum time evolved for the binaries used
 in a particular comparison. We start the window shortly after $t_{\rm mer}$ in order to exclude the turbulent merger phase from the resulting spectra. For example, the $R_{1.4} \simeq 11$~km, $L=40$~MeV
 EoS collapsed after $\sim15$~ms. Thus, in the following spectral comparisons of the
  $R_{1.4}\simeq 11$~km EoSs, we window both EoSs to the $\sim$14~ms window following merger,
  to ensure they have the same spectral resolution. For all other EoSs, the remnants do not
  collapse until the end of the evolution, so the windows are typically $\sim24$~ms.
  We then compute the Fourier transform of the windowed $\psi_4(t)$, using
  Welch's method with 8 overlapping segments for the longer-duration
  signals, and 7 overlapping segments for the $R_{1.4}\simeq 11$~km EoSs. Each
  segment is windowed with a Hann window and zero-padded to contain a total of
  4,096 points. From $\tilde{\psi}_{4}(f)$, we then calculate
$\tilde{h}(f)$, using the FFI technique of
\cite{Reisswig:2010di}. In this paper, we calculate $h_c(f)$ including all
$\ell=2,3$ modes.
 
Finally, we calculate the distinguishability of the post-merger GW signals using the overlap integral \cite{Lindblom:2008cm,McWilliams:2010eq}, defined as
\begin{equation}
O(h_1, h_2) = \frac{ \left<h_1, h_2\right>}{\sqrt{\left<h_1,h_1\right>\left<h_2, h_2\right>}},
\end{equation}
where $\left<h_i,h_j\right>$ is the inner product of two waveforms, given by
\begin{equation}
\left<h_i,h_j\right> = 4 \mathrm{Re} \int_{f_{\rm min}}^{f_{\rm max}} \frac{ \tilde{h}_i(f)  \tilde{h}_j^*(f) }{ S_n(f)} df,
\end{equation}
$S_n(f)$ is the power spectral noise of the detector, and $^*$ indicates the complex conjugate. For the noise curve, we use the design sensitivity curve for Advanced LIGO \cite{LIGOScientific:2014pky}, bounded between frequencies $f_{\rm min}$=1000~Hz and $f_{\rm max}=5000$~Hz. Values of the overlap integral smaller than $1-1/(2 \rho^2)$ are distinguishable, where $\rho$ is the signal-to-noise ratio (SNR). We consider the threshold SNR for detectability to be 8, in which case the criteria for distinguishability is $O \lesssim 0.992$.

\bibliography{main}

\end{document}